\documentclass[aps, prl, superscriptaddress, preprintnumbers, floatfix,twocolumn,10pt]{revtex4-2}

\usepackage{graphicx,amsfonts,amsmath,amssymb,amstext}
\usepackage{float,wrapfig}
\usepackage{subfigure, psfrag}
\usepackage{dsfont}
\usepackage{color}
\usepackage[colorlinks=true,urlcolor=blue,citecolor=blue,linkcolor=blue]{hyperref}
\usepackage{bm}

\newcommand{\be}{\begin{equation}}
\newcommand{\ee}{\end{equation}}
\newcommand{\ba}{\begin{eqnarray}}
\newcommand{\ea}{\end{eqnarray}}

\definecolor{purple}{rgb}{0.8,0,0.6}
\definecolor{darkgreen}{rgb}{0.00,0.6,0.00}

\extrafloats{100}

\begin{document}

\title{Strong suppression of electron convection in Dirac and Weyl semimetals}
\date{September 28, 2021}

\author{P.~O.~Sukhachov}
\email{pavlo.sukhachov@yale.edu}
\affiliation{Department of Physics, Yale University, New Haven, Connecticut 06520, USA}

\author{E. V. Gorbar}
\affiliation{Department of Physics, Taras Shevchenko National Kyiv University, Kyiv, 03022, Ukraine}
\affiliation{Bogolyubov Institute for Theoretical Physics, Kyiv, 03143, Ukraine}

\author{I.~A.~Shovkovy}
\affiliation{College of Integrative Sciences and Arts, Arizona State University, Mesa, Arizona 85212, USA}
\affiliation{Department of Physics, Arizona State University, Tempe, Arizona 85287, USA}

\begin{abstract}
It is shown that the convective instability in electron fluids in three- and two-dimensional (3D and 2D) Dirac and Weyl semimetals is strongly inhibited. The major obstacles for electron convection are the effects of the Coulomb forces and the momentum relaxation related to the interaction with impurities and phonons. The effect of the Coulomb forces is less pronounced in 2D materials, such as graphene. However, momentum relaxation still noticeably inhibits convection making it very difficult to achieve in practice.
\end{abstract}

\maketitle

{\it Introduction.---}
Electron hydrodynamics is an unusual transport regime that can be realized in clean crystals. As first conjectured by Gurzhi in the 1960s~\cite{Gurzhi:1963,Gurzhi:1968}, electrons could form a hydrodynamic fluid when the electron-electron scattering rate is larger than the scattering rates of electrons on impurities and phonons. In such a regime, the electron transport should reveal some conventional hydrodynamic effects, including the Poiseuille-like profile of the current in a wire, the formation of vortices, etc.

Historically, a hydrodynamic electron flow was first observed in a two-dimensional (2D) electron gas of high-mobility $\mathrm{(Al,Ga)As}$ heterostructures~\cite{Molenkamp-Jong:1994,Jong-Molenkamp:1995}. Later, a similar regime was confirmed in the ultrapure 2D metal palladium cobaltate (PdCoO$_2$)~\cite{Moll-Mackenzie:2016} and graphene~\cite{Crossno-Fong:2016,Ghahari-Kim:2016,Krishna-Falkovich:2017,Berdyugin-Bandurin:2018,Bandurin-Falkovich:2018,Ku-Walsworth:2019,Sulpizio-Ilani:2019}. (For reviews on electron hydrodynamics, see Refs.~\cite{Lucas-Fong:rev-2017,Narozhny:rev-2019}.) Electron hydrodynamics in graphene could be experimentally revealed via a negative nonlocal resistance and the formation of current vortices~\cite{Torre-Polini:2015,Pellegrino-Polini:2016,Levitov-Falkovich:2016,Falkovich-Levitov:2017,Danz-Narozhny:2019}, higher than ballistic conduction in constrictions~\cite{Guo-Levitov:2017,Krishna-Falkovich:2017}, and certain collective modes~\cite{Tomadin-Polini:2013,Svintsov-Otsuji:2013,Svintsov:2019,Mendl-Polini:2019}.
The profile of electric currents in the hydrodynamic regime can be reconstructed from the stray magnetic fields~\cite{Ku-Walsworth:2019} or the Hall field across the graphene ribbon~\cite{Sulpizio-Ilani:2019}. Recently, evidence of three-dimensional (3D) relativisticlike hydrodynamic electron transport was reported in the Weyl semimetal tungsten diphosphide WP$_2$~\cite{Gooth-Felser:2018}. This shows that Dirac and Weyl semimetals provide another promising platform for investigating the hydrodynamic regime of the electron transport in solids.

Dirac and Weyl semimetals are novel materials whose electron quasiparticle spectrum is described by the corresponding equations in the vicinity of the band-crossing points known as Dirac points and Weyl nodes~\cite{Katsnelson:book-2012,Armitage:rev-2018,GMSS:book}, respectively. Representative material realizations of Dirac semimetals include $A_3$Bi ($A=$ Na, K, Rb)~\cite{Liu-Chen-Na3Bi:2014,Xu-Hasan-Na3Bi:2015} and Cd$_3$As$_2$~\cite{Liu-Chen-Cd3As2:2014,Neupane-Hasan-Cd3As2:2014,Borisenko:2014} in 3D, as well as graphene in 2D. Weyl semimetals are realized, e.g., in transition metal monopnictides (TaAs, TaP, NbAs, and NbP)~\cite{Xu-Hasan:2015,Xu-Hasan-TaP:2015,Lv-Ding-TaAs:2015,Lv-Ding-TaAs:2015b,Yang-Chen-TaAs:2015,Xu-Shi-TaP:2015,Xu-Hasan-NbAs:2015}, EuCd$_2$As$_2$~\cite{Soh-Boothroyd:2019,Ma-Shi:2019}, Co$_3$Sn$_2$S$_2$~\cite{Liu-Felser-Co3Sn2S2:2018,Xu-Sun-Co3Sn2S2:2018,Wang-Lei-Co3Sn2S2:2018}, etc.

Guided by experience with conventional fluids, one may expect similar hydrodynamic effects to show up in the electron fluid. For example, the possibility of a preturbulent regime in graphene was proposed in Refs.~\cite{Mendoza-Succi:2011,Gabbana-Succi:2018}. (Kagome metals may provide a compelling platform for realizations of turbulence in electron fluids~\cite{Sante-Wehling:2020}.) The formation of the Rayleigh-B\'{e}nard convective cells in graphene was suggested and numerically studied in Ref.~\cite{Furtmaier-Herrmann:2015}. Generally, the Rayleigh-B\'{e}nard instability occurs in fluids subject to a buoyancy force (e.g., caused by gravity) and temperature gradient that results in a local thermal expansion of fluid~\cite{Benard:1900,Rayleigh:1916}. If the buoyancy force is strong enough, it becomes favorable to develop regular convective cells where the heat transfer is greatly assisted by the fluid motion. The observation of such cells would be a definitive signature of electron hydrodynamics. The possibility of electron convection is not only of academic interest but also can have important practical applications. Indeed, the convective regime may be invaluable for an effective heat transfer because the Nusselt number, which quantifies the ratio of convective to conductive heat transfer, is up to 10 (1000) times larger for a laminar (turbulent) flow. Since there are significant differences between conventional fluids and electron plasma in semimetals, it is necessary to investigate the onset conditions for electron convection in detail.

In this Letter, we show that the Rayleigh-B\'{e}nard instability in electron fluids in Dirac and Weyl semimetals is \emph{strongly inhibited}. We identify two major obstacles for the electron fluid convection: (i) Coulomb forces and (ii) momentum-relaxation effects. Unlike many conventional fluids, the electron plasma is electrically charged. In a semimetal, the overall electric neutrality is preserved by the compensating charge of immovable lattice ions. Any deviation from local neutrality induces a strong electric field and, consequently, is energetically unfavorable. In other words, the effects of the Coulomb forces strongly suppress local expansion and compression of the electron fluid. For the same reason, a temperature gradient cannot easily induce sufficiently large local density deviations needed for triggering convection.

We support this qualitative argument by a quantitative estimate of the Rayleigh number. Due to the aforementioned obstacles, the latter is so large in 3D Dirac and Weyl semimetals that achieving convection becomes practically impossible under any realistic conditions. As one might expect, the effects of the Coulomb forces are less pronounced in 2D materials. However, due to ubiquitous disorder effects and interaction with phonons, the momentum relaxation also noticeably  inhibits a convective flow in both 2D and 3D systems.

{\it Model.---}
The starting point in our discussion of the electron fluid is the Navier-Stokes equation~\cite{Lucas-Fong:rev-2017,Narozhny:rev-2019},
\begin{eqnarray}
\label{model-simple-Hydro-Weyl-NS}
&&\frac{1}{v_F^2}\left[\partial_t(\mathbf{u} w)  +\mathbf{u} w (\bm{\nabla}\cdot \mathbf{u})+(\mathbf{u}\cdot\bm{\nabla})(\mathbf{u} w) \right] -\eta \Delta \mathbf{u}
\nonumber\\
&& -\left[\zeta+\frac{\eta (d-2)}{d}\right] \bm{\nabla}\left(\bm{\nabla}\cdot\mathbf{u}\right)
=-\bm{\nabla}P -\frac{w \mathbf{u}}{v_F^2\tau} -en\mathbf{E}.\nonumber\\
\end{eqnarray}
Here, $w=\epsilon+P$ is the enthalpy, $\epsilon$ is the energy density, $P$ is the pressure, $\mathbf{u}$ is the electron fluid velocity, $n$ is the electron number density, $-e$ is the electron charge, and $v_F$ is the Fermi velocity.
Further, $\eta$ and $\zeta$ are the shear and bulk viscosities, respectively. In relativisticlike systems, $\zeta\approx0$~\footnote{Not only is the bulk viscosity negligible, but it would not play any significant role for the convective instability because the Coulomb forces strongly suppress compressions and expansions of the electron fluid.} and $\eta = \eta_{\rm kin}w/v_F^2$, where $\eta_{\rm kin}\sim v_F^2\tau_{\rm ee}$ is the kinematic shear viscosity and $\tau_{\rm ee}$ is the electron-electron interaction time (see, e.g., Refs.~\cite{Principi-Polini:2015,Alekseev:2016}). For a relativisticlike fluid, $P=\epsilon/d$ and $w=(d+1)\epsilon/d$, where $d$ is the spatial dimension. The momentum relaxation, which is inevitable in real solids, is quantified by the relaxation time $\tau$ that contains contributions from scattering on impurities and phonons. For simplicity, we assume that the electron fluid is isotropic, which is sufficient for the purposes of this study. Finally, we account for an electric field $\mathbf{E}$, which includes both external and induced contributions.

The electric and energy current densities are
\begin{eqnarray}
\label{convection-model-J-def}
\mathbf{J} &=& -en\mathbf{u} +\sigma \left[\mathbf{E} +\frac{T}{e}\bm{\nabla}\left(\frac{\mu}{T}\right)\right],\\
\label{convection-model-J-eps-def}
\mathbf{J}^{\epsilon} &=& w\mathbf{u}  -\eta \left[\left(\bm{\nabla}\cdot\mathbf{u}\right) \mathbf{u} +u_{j}\bm{\nabla}u_{j} -\frac{2}{d} \mathbf{u} \left(\bm{\nabla}\cdot\mathbf{u}\right)\right].
\end{eqnarray}
Here, $\mu$ is the chemical potential, $T$ is temperature, and $\sigma$ is the intrinsic conductivity~\cite{Lucas-Fong:rev-2017}.
The electric and energy currents satisfy the standard continuity relations
\begin{equation}
\label{model-simple-electric-current-div}
-e\partial_t n + (\bm{\nabla}\cdot \mathbf{J}) = 0
\end{equation}
and
\begin{equation}
\label{model-simple-Hydro-WeylD -energy}
\partial_t \epsilon  +\left(\bm{\nabla}\cdot\mathbf{J}^{\epsilon}\right)=\left(\mathbf{E}\cdot\mathbf{J}\right),
\end{equation}
respectively.

Since the electron fluid is electrically charged, the hydrodynamic equations should be supplemented by Maxwell's equations. By assuming a slow flow, we neglect the effects of dynamic magnetic fields on fluid motion. In such a quasistatic approximation, only the Gauss law
\begin{equation}
\label{model-Hydro-Weyl-Maxwell-be}
\bm{\nabla}\cdot\mathbf{E} = -4\pi e \delta n
\end{equation}
is relevant. Here, $-e\delta n$ is the deviation of the electron charge density from the background equilibrium value. As we will show below, electric fields induced by $\delta n$ play a profound role in suppressing electron convection in 3D. The Coulomb forces also hinder convection in 2D systems, but their effect is less dramatic.

It is worth noting that the Coulomb interactions are responsible for both viscosity of the electron fluid and screening effects. Indeed, it is the microscopic interparticle Coulomb force that governs the electron-electron scattering and the formation of electron fluid. On the other hand, the Gauss law in Eq.~(\ref{model-Hydro-Weyl-Maxwell-be}) determines a background electric field, which comes as an average uncompensated field over macroscopic distances. This field is induced when the electron fluid is compressed or expanded locally with respect to the ion lattice.

To investigate the possibility of electron convection, we follow the same conceptual approach as in conventional fluids (see, e.g., Refs.~\cite{Landau:t6-2013,Chandrasekhar:book}), but amend the hydrodynamic equations with the Gauss law. As the first step, we find the steady-state solutions for the temperature profile and the electric field in the absence of hydrodynamic flow. Then, by using these solutions as a background, we derive the threshold criterion for convection.

{\it Steady-state solution without flow.---}
Let us start by determining the steady-state solution for the electric field and temperature inside a slab of finite thickness $L$ along the $x$ direction. For simplicity, we assume that the slab is infinite along other directions.

Temperature $T(x)$ and the electric potential $\varphi(x)$ take different values on the opposite surfaces of the slab, i.e.,
\begin{eqnarray}
\label{sol-steady-T-BC}
T(x=0)=T_{\rm L}, \qquad T(x=L)=T_{\rm R},
\\
\label{sol-steady-static-phi-BC-1}
\varphi(x=0) = \varphi_{\rm L}, \qquad \varphi(x=L ) = \varphi_{\rm R}.
\end{eqnarray}
We employ a perturbation scheme where deviations of all quantities (denoted by a tilde) are small compared to their global equilibrium values (denoted by subscript $0$), e.g., $\tilde{T}/T_0\ll1$, etc. By setting $\mathbf{u}=\mathbf{0}$ in Eqs.~(\ref{model-simple-Hydro-Weyl-NS}) through (\ref{model-Hydro-Weyl-Maxwell-be}), one finds that the temperature function $\tilde{T}(x)$ is determined by the Laplace equation, $\Delta \tilde{T}=0$. Its solution that satisfies the boundary conditions (\ref{sol-steady-T-BC}) reads
\begin{equation}
\label{convection-steady-tT-sol}
\tilde{T} = T_{\rm L}-T_0 + \frac{T_{\rm R}-T_{\rm L}}{L }x.
\end{equation}
By taking into account this solution, Eqs.~(\ref{model-simple-electric-current-div}) and (\ref{model-Hydro-Weyl-Maxwell-be}) can be rewritten as follows:
\begin{eqnarray}
\label{convection-steady-mu-eq}
\Delta \tilde{\mu} &=& 4\pi e^2\tilde{n}
= q_{\rm TF}^2\tilde{\mu} +4\pi e^2(\partial_{T} n) \tilde{T},\\
\label{convection-steady-phi-eq}
\Delta \tilde{\varphi} &=& -4\pi e\tilde{n},
\end{eqnarray}
where $q_{\rm TF} = \sqrt{4\pi e^2(\partial_{\mu} n)}$ is the Thomas-Fermi wave vector. Its explicit expression is given in the Supplemental Material~\cite{SM}.

By using the equation of motion (\ref{model-simple-Hydro-Weyl-NS}) at $\mathbf{u}=\mathbf{0}$ and requiring the overall charge neutrality of the sample, we obtain the steady-state expressions for $\tilde{\mu}$ and $\tilde{\varphi}$ (see Supplemental Material~\cite{SM}). They reveal that the electric field is nonvanishing but screened strongly inside the sample. Also, the electric charge deviations are small ($\tilde{n}\approx0$) in the bulk of the slab. However, both electric field and charge density are noticeably nonuniform near the surfaces. Since convection is determined largely by bulk properties, it is justified to ignore the inhomogeneity of the electric field and other variables near the surfaces. Such an approximation simplifies analytical calculations but should not affect the qualitative results for sufficiently thick samples.

{\it Convection threshold.---}
To determine the threshold of convection, we consider a hydrodynamic flow in the background of the screened electric field and the temperature gradient obtained in the previous section. As in conventional fluids~\cite{Landau:t6-2013,Chandrasekhar:book}, the analysis is simplified by using an analog of the Oberbeck-Boussinesq approximation. The resulting system, which is linear in the flow velocity $\mathbf{u}$ and the flow-driven deviations $P_u$, $\mathbf{E}_u$, $n_u$, $\mu_u$, and $T_u$ (appearing on top of the steady-state solution), reads
\begin{eqnarray}
\label{sol-steady-u-Hydro-Weyl-NS-1}
&&\bm{\nabla}P_u -\eta \Delta \mathbf{u} -\left[\zeta+\frac{\eta (d-2)}{d}\right] \bm{\nabla}\left(\bm{\nabla}\cdot\mathbf{u}\right) \nonumber\\
&&\hspace{1.25in}= -\frac{w_0 \mathbf{u}}{v_F^2\tau} -en_0\mathbf{E}_u -en_u\tilde{\mathbf{E}},\\
\label{sol-steady-u-Hydro-Weyl-energy-1}
&&(\mathbf{u}\cdot\bm{\nabla})\tilde{w}+w_0(\bm{\nabla}\cdot\mathbf{u}) =0,\\
\label{sol-steady-u-eqs-cont-1}
&&-en_0(\bm{\nabla}\cdot\mathbf{u})+\sigma(\bm{\nabla}\cdot \mathbf{E}_u) +\frac{\sigma}{e}\left(\Delta \mu_u -\frac{\mu_0}{T_0} \Delta T_u \right)=0,\nonumber\\
\\
\label{sol-steady-u-eqs-Gauss-1}
&&\bm{\nabla}\cdot \mathbf{E}_u = -4\pi e n_u.
\end{eqnarray}
Here, we assumed that the cross-terms containing fluid flow velocity and the temperature gradient are small compared to other terms. In addition, we neglected the terms that are of the second order in steady-state deviations (e.g., $\tilde{n}\tilde{\mathbf{E}}\ll n_0\tilde{\mathbf{E}}$).

The last term on the right-hand side in Eq.~(\ref{sol-steady-u-Hydro-Weyl-NS-1}) and the first term in Eq.~(\ref{sol-steady-u-Hydro-Weyl-energy-1}) are crucial for driving convection and their analogs are included in the Oberbeck-Boussinesq approximation for conventional fluids. Indeed, the term $-en_u\tilde{\mathbf{E}}$ is similar to the buoyancy force for regular fluids. Recall that buoyancy is the consequence of an external (e.g., gravitational) force exerted on a fluid with a density gradient, typically induced by a temperature difference between the top and bottom surfaces.
In the case of electrons, gravitation has negligible effects and the role of buoyancy force is played by the electric force related to the in-medium electric field $\tilde{\mathbf{E}}$. Therefore, another key requirement for achieving convection is the nonzero compressibility of the fluid. It is taken into account by the first term in Eq.~(\ref{sol-steady-u-Hydro-Weyl-energy-1}).

It is very important for the problem under consideration that, unlike ordinary fluids, expansion and compression of the electron fluid give rise to strong electric fields. Indeed, in view of the Gauss law (\ref{sol-steady-u-eqs-Gauss-1}), any change of the electric charge density leads to an electric field. As we will explicitly show below, this is one of the key factors inhibiting convection in electron fluids as the energy price for the appearance of electric fields is very high.

To estimate the threshold for convection, we use the bulk steady-state solutions, $\tilde{n}\approx0$ and $\tilde{\mathbf{E}}=\tilde{E} \hat{\mathbf{x}}$, which include both external and induced fields. Then, $(\mathbf{u}\cdot\bm{\nabla})\tilde{w} = u_x (\partial_x\tilde{w}) \propto u_x \left(T_{R}-T_{L}\right)$. The explicit expression for $(\partial_x\tilde{w})$ as well as the general solutions to Eqs.~(\ref{sol-steady-u-Hydro-Weyl-NS-1}) through (\ref{sol-steady-u-eqs-Gauss-1}) are given in the Supplemental Material~\cite{SM}.

We use a plane-wave ansatz for the hydrodynamic variables $u_x$, $T_u$, and $\mu_u$, e.g.,
\begin{equation}
\label{Solutions-steady-hydro-Tu}
T_u = C_T e^{i\mathbf{k}_{\perp}\cdot\mathbf{r}_{\perp}} e^{ik_xx},
\end{equation}
where $\mathbf{k}_{\perp}$ is the wave vector perpendicular to the surface normal.
The characteristic equation for the system of Eqs.~(\ref{sol-steady-u-Hydro-Weyl-NS-1}) through (\ref{sol-steady-u-eqs-Gauss-1}) reads~\cite{SM}
\begin{equation}
\label{Solutions-steady-hydro-char-eq-3}
k^2\left(k^2+\frac{1}{\lambda_{G}^2}\right)\left(k^2 +q_{\rm TF}^2\right) -\frac{k_{\perp}^2}{L ^4} \mathrm{Ra} =0,
\end{equation}
where $k^2=k_{\perp}^2+k_x^2$. In Eq.~(\ref{Solutions-steady-hydro-char-eq-3}), we used the following shorthand notations:
\begin{eqnarray}
\label{Solutions-steady-hydro-Rtau-def}
\lambda_{G} &=& \sqrt{\frac{v_F^2\tau \eta}{w_0}} = \sqrt{\tau \eta_{\rm kin}},\\
\label{Solutions-steady-hydro-RQ-def}
\mathrm{Ra} &=& L^4\frac{e^3n_0 \tilde{E} (\partial_x \tilde{w}) T_0}{\sigma w_0^2\eta}  \left[n_0(\partial_{T}n) -s_0(\partial_{\mu}n)\right].
\end{eqnarray}
Here, $\lambda_{G}$ is the Gurzhi length that quantifies the momentum relaxation and $\mathrm{Ra}$ is the Rayleigh number. In the limit $\lambda_{G}\to\infty$ and $q_{\rm TF}\to0$, the characteristic equation (\ref{Solutions-steady-hydro-char-eq-3}) coincides with the textbook result for conventional fluids (cf. Ref.~\cite{Landau:t6-2013}). Convective instability characterized by periodic spatial pattern of the fluid velocity is realized for real $k_x$ and $k_{\perp}$. This requires
$\mathrm{Ra}\gtrsim \mathrm{Ra}_{\rm min}$, where $\mathrm{Ra}_{\rm min}$ is determined from the characteristic equation (\ref{Solutions-steady-hydro-char-eq-3}) with wave vectors constrained by boundary conditions.

To derive the convection threshold $\mathrm{Ra}_{\rm min}$, let us determine the allowed values of $k_x$ in Eq.~(\ref{Solutions-steady-hydro-char-eq-3}). They follow from the boundary conditions for $T_u$ and $\mathbf{u}$, i.e.,
\begin{equation}
\label{Solutions-steady-hydro-BC-1}
T_u(x=0,L )=0, \quad u_x(x=0,L )=0.
\end{equation}
For the perpendicular components of velocity $\mathbf{u}_{\perp}$, we employ the free-surface boundary conditions, which provide the most conservative estimate of the convection threshold. Indeed, the no-slip boundary conditions can be considered as a source of additional dissipation that further inhibits convection; see, e.g., Ref.~\cite{Chandrasekhar:book} for neutral fluids. We found that the free-surface boundary conditions are satisfied for $|k_x|=\pi n/L $ with $n=1,2,3,\ldots$ (see Supplemental Material~\cite{SM} for details). It is worth noting, however, that the exact form of the boundary conditions is not important for our qualitative arguments because they only determine the allowed values of $k_x$.

The characteristic equation (\ref{Solutions-steady-hydro-char-eq-3}) gives the following relation between the Rayleigh number and wave vector:
\begin{equation}
\label{Solutions-steady-hydro-RQ-sol}
\mathrm{Ra} = L ^4\frac{\left(k_{\perp}^2+k_x^2\right)\left(k_{\perp}^2+k_x^2+\lambda_{G}^{-2}\right)\left(k_{\perp}^2+k_x^2+q_{\rm TF}^2\right) }{k_{\perp}^2}.
\end{equation}
This shows that the Coulomb forces and the momentum relaxation effects, quantified by $q_{\rm TF}$ and $\lambda_{G}$, respectively, increase the minimal value of the Rayleigh number $\mathrm{Ra}_{\rm min}$ needed to achieve convection.
In order to determine $\mathrm{Ra}_{\rm min}$, one should minimize Eq.~(\ref{Solutions-steady-hydro-RQ-sol}) with respect to the wave vectors allowed by the boundary conditions, i.e., at $k_x=\pi/L$. The general expression for the minimal Rayleigh number at finite $\lambda_{G}$ and nonzero $q_{\rm TF}$ is cumbersome (see also Supplemental Material~\cite{SM}). Since the Rayleigh number $\mathrm{Ra}$ increases with $k_x$, we can estimate the corresponding lower bound $\mathrm{Ra}_{0}$ (which is smaller than the actual $\mathrm{Ra}_{\rm min}$) by setting $k_x=0$ and neglecting $k_{\perp}$ compared to $q_{\rm TF}$ and $\lambda_{G}^{-1}$ in Eq.~(\ref{Solutions-steady-hydro-RQ-sol}), i.e.,
\begin{equation}
\label{Solutions-steady-hydro-RQ-min-LB}
\mathrm{Ra}_{0}=L^4q_{\rm TF}^2/\lambda_{G}^2 < \mathrm{Ra}_{\rm min}.
\end{equation}
It is instructive to mention that $q_{\rm TF}\gg\pi/L $ and $\lambda_{G}\ll L /\pi$ hold for realistic samples of 3D Dirac and Weyl semimetals and, therefore, the above estimate is indeed reasonable. To verify this, let us consider typical parameters for semimetals, e.g., $\mu_0=20~\mbox{meV}$, $T_0=25~\mbox{K}$, and use the Fermi velocity $v_F\approx1.4\times10^7~\mbox{cm/s}$~\cite{Kumar-Felser:2017}. In this case, one obtains $q_{\rm TF}\approx 9.9\times10^{6}~\mbox{cm}^{-1}$ by assuming quasiparticles with a 3D relativisticlike energy spectrum. For such a large Thomas-Fermi wave vector, one finds that $q_{\rm TF}\gg\pi/L $ in a wide range of experimentally achievable samples~\footnote{One might naively suggest that convection could be still observed in thin films $L \ll \pi/q_{\rm TF}$. However, a small thickness of the slab makes external forces and thermal gradients less efficient [see Eqs.~(\ref{Solutions-steady-hydro-RQ-def}) and (\ref{Solutions-steady-hydro-RQ-actual})].}. As for the momentum relaxation length, we estimate $\lambda_{G}\approx 0.4~\mu\mbox{m}$ at $\tau=0.1~\mbox{ns}$ and $\tau_{\rm ee}=0.3~\mbox{ps}$, which is quite small compared to a typical thickness of Dirac semimetal slabs.

By using Eq.~(\ref{Solutions-steady-hydro-RQ-min-LB}) and the same characteristic parameters as before, we estimate the lower bound for $\mathrm{Ra}_{\rm min}$ as
\begin{equation}
\label{Solutions-steady-hydro-RQ-min-estim-5}
\mathrm{Ra}_{0}=L^4q_{\rm TF}^2/\lambda_{G}^2\approx 6.5\times10^{22}\, 
\left(\frac{L}{1~\mbox{cm}}\right)^4.
\end{equation}
It is instructive to compare this estimate with the benchmark result for conventional fluids. By taking $\lambda_{G}\to\infty$ and $q_{\rm TF}=0$, we find that the critical value of the Rayleigh number is reached for $k_x=\pi/L$ and $k_{\perp,\rm min}=\pi/(\sqrt{2}L )$ giving $\mathrm{Ra}_{\rm min}=27\pi^4/4 \approx 657.5$.

The enormous value of the lower bound for the minimal Rayleigh number~(\ref{Solutions-steady-hydro-RQ-min-estim-5}) implies that convection in the electron fluid is strongly suppressed. Indeed, Eq.~(\ref{Solutions-steady-hydro-RQ-def}) gives the following estimate for the Rayleigh number in Dirac semimetals:
\begin{equation}
\label{Solutions-steady-hydro-RQ-actual}
\mathrm{Ra} \approx 4\times10^{9}\,\frac{\delta T}{T_0} 
\frac{\tilde{E}}{1~\mbox{V/m}} \left(\frac{L}{1~\mbox{cm}}\right)^3,
\end{equation}
where $\tilde{E}$ is the in-medium field that includes both external and induced components, and $\delta T/T_0$ is the normalized temperature difference between the slab surfaces. As is easy to see, the estimate in Eq.~(\ref{Solutions-steady-hydro-RQ-actual}) is many orders of magnitude smaller than $\mathrm{Ra}_{0}$ for any reasonable electric field and temperature gradient available in experiments. Thus, in agreement with the heuristic arguments, convection is ruled out for 3D Dirac semimetals.

\begin{figure}[t]
\includegraphics[width=0.4\textwidth]{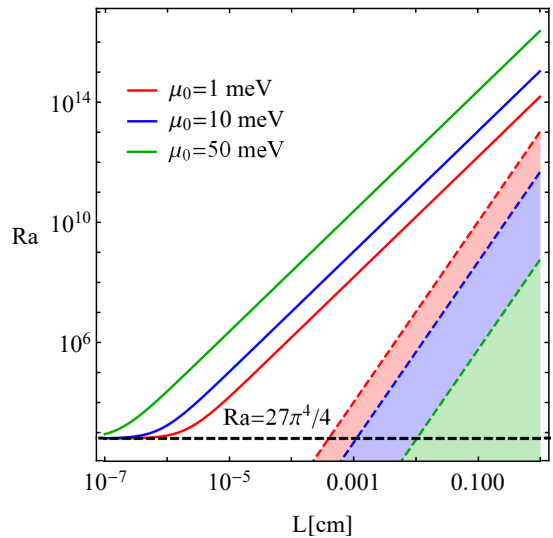}
\caption{Solid lines show the minimal Rayleigh number needed for convection. The realistic Rayleigh numbers achievable in Dirac semimetals are shown by the shaded regions. Dashed lines show the Rayleigh number
(\ref{Solutions-steady-hydro-RQ-actual}) calculated for $\delta T=T_0$ and $\tilde{E}=10~\mbox{V/m}$.
The Gurzhi length is $\lambda_{G}\approx0.4~\mu\mbox{m}$ and the values of the Thomas-Fermi wave vector are $q_{\rm TF}\approx 2\times10^{6}~\mbox{cm}^{-1}$ at $\mu_0=1~\mbox{meV}$, $q_{\rm TF}\approx 5.2\times10^{6}~\mbox{cm}^{-1}$ at $\mu_0=10~\mbox{meV}$, and $q_{\rm TF}\approx 2.4\times10^{7}~\mbox{cm}^{-1}$ at $\mu_0=50~\mbox{meV}$.
}
\label{fig:convection-Ra}
\end{figure}

The minimal Rayleigh number $\mathrm{Ra}_{\rm min}$ as a function of the slab width $L$ is presented by solid lines in Fig.~\ref{fig:convection-Ra} for the three fixed values of the chemical potential: $\mu_{0}=1~\mbox{meV}$ (red), $\mu_{0}=10~\mbox{meV}$ (blue), and $\mu_{0}=50~\mbox{meV}$ (green)~\footnote{To accommodate even extreme possibilities, we extend the range of the Fermi energies down to the unrealistically small $\mu_{0}=1~\mbox{meV}$.}.
As we see, the minimal Rayleigh number needed for convection is enormous for macroscopic samples. It approaches the value in conventional fluids, i.e., $\mathrm{Ra}_{\rm min}=27\pi^4/4$, only for extremely thin slabs. In the same figure, the shaded regions show the ranges of realistic estimates for the Rayleigh number obtained from Eq.~(\ref{Solutions-steady-hydro-RQ-actual}) by assuming rather conservative values $\delta T\leq T_0$ and $\tilde{E}\leq 10~\mbox{V/m}$. It is clear that the realistic values of the Rayleigh number are many orders of magnitude smaller than $\mathrm{Ra}_{\rm min}$ required for convection.

{\it Graphene.---}
In view of the great interest in electron hydrodynamics in graphene, let us re-examine whether the convective instability is possible in the 2D case. While the electron convection in graphene was already studied in Ref.~\cite{Furtmaier-Herrmann:2015}, the effects of the Coulomb forces and the momentum dissipation were not taken into account. The screening effects in gated graphene can be treated in the ``gradual channel" approximation~\cite{Shur:book,Dyakonov-Shur:1993} (see also Ref.~\cite{Tomadin-Polini:2013}). In this approximation, the induced electric field is given by
\begin{equation}
\label{Solutions-graphene-capacitance-hydro}
\mathbf{E}_u=  \frac{e}{C} \bm{\nabla} n_u,
\end{equation}
where $C=\varepsilon/(4\pi L_g)$ is the capacitance per unit area, $\varepsilon$ is the dielectric constant, and $L_g$ is the distance to the gate. The minimum Rayleigh number is determined similarly to the 3D case considered above, but the Gauss law (\ref{sol-steady-u-eqs-Gauss-1}) is replaced by Eq.~(\ref{Solutions-graphene-capacitance-hydro}) (for details, see Supplemental Material~\cite{SM}). We obtain
\begin{equation}
\label{Solutions-graphene-hydro-RQ-sol}
\mathrm{Ra} = L ^4\frac{\left(k_{\perp}^2+k_x^2\right)^2\left(k_{\perp}^2+k_x^2+\lambda_{G}^{-2}\right)\left(1+Q^2\right)}{k_{\perp}^2},
\end{equation}
where $Q=\sqrt{e^2 (\partial_{\mu}n)/C}$. For $T_u\propto \sin{(k_xx)}$, the boundary conditions are satisfied for $|k_x|=\pi n/L $ and $n=1,2,3,\ldots$. As in the 3D case, the Coulomb forces and momentum relaxation increase the minimal Rayleigh number needed to achieve convection.

Quantitatively, the effects of the Coulomb forces are much weaker in 2D. This is also supported by numerical estimates. Indeed, by using typical parameters for graphene, i.e., $v_F=1.1\times10^{8}~\mbox{cm/s}$, $\mu_0=100~\mbox{meV}$, $T_0=100~\mbox{K}$, $L_{g}=100~\mbox{nm}$, and $\varepsilon=1$, we estimate $Q\approx 6.6$ (see Supplemental Material~\cite{SM} for details). Therefore, the Coulomb forces increase $\mathrm{Ra}_{\rm min}$ only by about an order of magnitude. The momentum relaxation effects, on the other hand, are very important. We estimate $\lambda_{G}\approx 2.6~\mu\mbox{m}$ at $\tau=0.1~\mbox{ns}$ and $\tau_{\rm ee}=(\hbar^2 v_F^2/e^4) (\hbar \mu_0/T_0^2)\approx 0.2~\mbox{ps}$. The corresponding minimal value of the Rayleigh number is
\begin{equation}
\label{Solutions-graphene-hydro-RQ-min-estim-1}
\mathrm{Ra}_{\rm min} \approx 2.7\times10^{6}
\end{equation}
at $L =100~\mu\mbox{m}$. This is almost four orders of magnitude larger than the benchmark value $27\pi^4/4$.
The Rayleigh number for graphene is estimated as
\begin{equation}
\label{Solutions-steady-hydro-RQ-graphene-actual}
\mathrm{Ra} \approx 4.4\times10^{7} \,\frac{\delta T}{T_0}
\frac{\tilde{E}}{1~\mbox{V/m}} \left(\frac{L}{1~\mbox{cm}}\right)^3,
\end{equation}
Our estimate suggests that in order to exceed the minimal Rayleigh number (\ref{Solutions-graphene-hydro-RQ-min-estim-1}), centimeter-sized graphene samples are needed when the total electric fields are of the order of $1~\mbox{V/m}$.

Thus, the momentum relaxation significantly inhibits convection in 2D systems too. The effect of the Coulomb forces, however, is less pronounced compared to the case of 3D Dirac semimetals. This situation resembles the role of the Coulomb forces in the spectrum of plasmons. Indeed, the plasmon dispersion relation is gapped in 3D due to the efficient screening of electric charge oscillations. On the other hand, in 2D systems the spectrum of plasmons remains gapless in the gradual channel approximation, where the Coulomb forces only enhance the plasma velocity.

{\it Summary.---} In this Letter, we showed that the convective instability is strongly inhibited in the electron fluid in 2D and 3D Dirac and Weyl semimetals. We identified the following two major inhibitors: (i) the Coulomb forces and (ii) the momentum relaxation effects due to scattering on impurities and phonons. Both are unavoidable and play a critical role in the charged electron fluid in semimetals.
For realistic parameters in 3D Dirac and Weyl semimetals, the effect of the Coulomb forces dominates over the momentum relaxation and leads to an extremely large convection threshold. In 2D systems such as graphene, the key role is played by the momentum relaxation that increases the minimum Rayleigh number needed for convection. The corresponding threshold values are a few orders of magnitude larger than in conventional fluids. Our findings imply that the electron fluid convection is unlikely to be realized in Dirac and Weyl semimetals.

While we focused on the fluid of electron quasiparticles with an isotropic relativisticlike dispersion relation, similar arguments should apply to systems with anisotropic and even nonrelativisticlike dispersion relations, although the quantitative details will be different. For example, the critical value of the Rayleigh number may be different along different crystal directions for materials with anisotropic spectra. Because of the crucial role of the Coulomb forces, we speculate that fluids made of neutral quasiparticles such as magnons (if their hydrodynamic regime is realized~\cite{Iacocca-Hoefer:2019,Ulloa-Duine:2019}) promise to be better candidates for convection in solid-state materials. It is tempting to suggest that phonon fluids~\cite{Gurzhi:1968,Nielsen-Shklovskii:1969,Cepellotti-Marzari:2015,Lee-Li:review-2019,Machida-Behnia:2019} might also demonstrate convective instability. It is not clear, however, whether the analogy with conventional fluids is complete and the equivalent of buoyancy forces is present.

\begin{acknowledgments}
P.O.S. acknowledges the support through the Yale Prize Postdoctoral Fellowship in Condensed Matter Theory.
E.V.G. acknowledges a collaboration within the Ukrainian-Israeli Scientific Research Program of the Ministry of Education and Science of Ukraine (MESU) and the Ministry of Science and Technology of the state of Israel (MOST). The work of I.A.S. was supported by the U.S. National Science Foundation under Grant No.~PHY-1713950.
\end{acknowledgments}

\bibliography{library-short}

\end{document}


\title{Supplemental Material\\
Strong suppression of electron convection in Dirac and Weyl semimetals}

\author{P.~O.~Sukhachov}
\email{pavlo.sukhachov@yale.edu}
\affiliation{Department of Physics, Yale University, New Haven, Connecticut 06520, USA}

\author{E. V. Gorbar}
\affiliation{Department of Physics, Taras Shevchenko National Kyiv University, Kyiv, 03022, Ukraine}
\affiliation{Bogolyubov Institute for Theoretical Physics, Kyiv, 03143, Ukraine}

\author{I.~A.~Shovkovy}
\affiliation{College of Integrative Sciences and Arts, Arizona State University, Mesa, Arizona 85212, USA}
\affiliation{Department of Physics, Arizona State University, Tempe, Arizona 85287, USA}

\maketitle
\tableofcontents

\section{S I. Thermodynamic variables}
\label{sec:app-def}

In this section, we present thermodynamic variables used in our analysis of electron convection. We employ a relativisticlike isotropic dispersion relation for the electron quasiparticles $\epsilon_{p}=v_Fp$ in both 2D and 3D systems.

In 3D, the electron number density $n$ equals
\begin{equation}
\label{app-model-n}
n= -N_g\frac{T^{3}}{\pi^2 \hbar^3 v_F^{3}}  \left[\mbox{Li}_{3}\left(-e^{\mu/T}\right) -\mbox{Li}_{3}\left(-e^{-\mu/T}\right)\right] = N_g\frac{\mu \left(\mu^2 +\pi^2T^2\right)}{6\pi^2 v_F^3 \hbar^3},
\end{equation}
where $\mbox{Li}_{j}(x)$ is the polylogarithm function. The energy density $\epsilon$ reads
\begin{equation}
\label{app-model-epsilon}
\epsilon =-N_g\frac{3T^{4}}{\pi^2 \hbar^3 v_F^{4}}  \left[\mbox{Li}_{4}\left(-e^{\mu/T}\right) +\mbox{Li}_{4}\left(-e^{-\mu/T}\right)\right]= N_g\frac{1}{8\pi^2 \hbar^3v_F^3} \left(\mu^4 +2\pi^2T^2\mu^2 +\frac{7\pi^4T^4}{15}\right),
\end{equation}
where the prefactor $N_g$ in Eqs.~(\ref{app-model-n-2D}) and (\ref{app-model-eps-2D}) is the degeneracy factor. We have $N_g=2$ in 3D Dirac semimetals with a single Dirac point and $N_g=4$, which accounts for the valley and spin degeneracy, in graphene. In addition, $v_F$ is the Fermi velocity, $\mu$ is the chemical potential, and $T$ is temperature.

The electron number density $n$ and the energy density $\epsilon$ for a 2D relativisticlike spectrum are
\begin{equation}
\label{app-model-n-2D}
n= -N_g \frac{T^2}{2\pi v_F^2 \hbar^2} \left[\mbox{Li}_{2}\left(-e^{\mu/T}\right) -\mbox{Li}_{2}\left(-e^{-\mu/T}\right)\right]
\end{equation}
and
\begin{equation}
\label{app-model-eps-2D}
\epsilon = -N_g \frac{T^3}{\pi v_F^2 \hbar^2} \left[\mbox{Li}_{3}\left(-e^{\mu/T}\right) +\mbox{Li}_{3}\left(-e^{-\mu/T}\right)\right].
\end{equation}

The pressure $P$ and the enthalpy density $w=\epsilon+P$ are given by the standard expressions for a relativisticlike systems, $P=\epsilon/d$ and $w=(d+1)\epsilon/d$, where $d=2,3$ is the spatial dimension. Note that the following relations for the derivatives with respect to chemical potential and temperature are valid:
\begin{equation}
\label{app-model-epsilon-mu}
\partial_{\mu}\epsilon = n d, \quad \partial_{T}\epsilon = s d,
\end{equation}
where $s=(w-\mu n)/T$ is the entropy density.

\section{S II. Steady state}
\label{app-sec:steady}

In this section, we discuss the temperature profile, the electric field, and the charge density for the steady state in the model defined in the main text. In particular, we consider a slab of a Dirac or Weyl semimetal with a finite thickness $L$ along the $x$ direction. For simplicity, we assume that the slab is infinite in other directions. In the absence of hydrodynamic flow, the local deviations of the chemical potential $\tilde{\mu}$ and temperature $\tilde{T}$ from their values $T_0$ and $\mu_0$ in global equilibrium are caused by an external electric field and a temperature gradient. In other words,
\begin{eqnarray}
\label{app-sol-steady-T-series}
T &\approx& T_0 + \tilde{T},\\
\label{app-sol-steady-mu-series}
\mu &\approx& \mu_0 + \tilde{\mu},\\
\label{app-sol-steady-E-series}
\mathbf{E} &\approx& \tilde{\mathbf{E}},
\end{eqnarray}
where, by assumption, the deviations $\tilde{\mu}$ and $\tilde{T}$ are small compared to the equilibrium values. As we will show below, the temperature gradient and the external electric field are included via the appropriate boundary conditions.

To the leading order in deviations, the equation of motion, the electric current continuity equation, and the Gauss law read
\begin{eqnarray}
\label{app-sol-steady-static-eqs-NS-1}
&&\bm{\nabla} \tilde{P} = -en_0 \tilde{\mathbf{E}},\\
\label{app-sol-steady-static-eqs-cont-1}
&&\sigma\left(\bm{\nabla}\cdot \tilde{\mathbf{E}}\right) +\frac{\sigma}{e}\Delta \left(\tilde{\mu} -\frac{\mu_0}{T_0} \tilde{T}\right) = 0,\\
\label{app-sol-steady-static-eqs-Gauss-1}
&&\bm{\nabla}\cdot \tilde{\mathbf{E}} = -4\pi e \tilde{n},
\end{eqnarray}
respectively. Here, $-e$ is the charge of the electron and $\sigma$ is the intrinsic conductivity~\cite{Lucas-Fong:rev-2017}. For the intrinsic conductivity, we use the result obtained in a holographic approach~\cite{Kovtun-Ritz:2008,Hartnoll:2014,Landsteiner-Sun:2015,Davison-Hartnoll:2015}, i.e.,
\begin{equation}
\label{app-sol-steady-static-sigma-def}
\sigma = \frac{3 \pi\hbar v_F^3}{2} (\partial_{\mu} n) \tau_{\rm ee},
\end{equation}
where $\tau_{\rm ee}$ is the electron-electron scattering time. The intrinsic conductivity in the hydrodynamic regime of graphene can be estimated as~\cite{Hartnoll-Sachdev:2007}
\begin{equation}
\label{app-Solutions-graphene-hydro-sigma-def}
\sigma = \frac{2e^2}{\pi \hbar}.
\end{equation}

By applying $\bm{\nabla}$ to Eq.~(\ref{app-sol-steady-static-eqs-NS-1}) and making use of Eqs.~(\ref{app-sol-steady-static-eqs-cont-1}) and (\ref{app-sol-steady-static-eqs-Gauss-1}), we obtain the following equation for temperature deviations:
\begin{equation}
\label{app-convection-steady-thermo-2}
\Delta \tilde{T}=0.
\end{equation}
In the derivation, we also took into account that
\begin{equation}
\label{app-sol-steady-static-nabla-P}
\bm{\nabla} \tilde{P} = n_0\bm{\nabla}\tilde{\mu} +s_0 \bm{\nabla} \tilde{T}
\end{equation}
and neglected terms of higher order in deviations.

In our setup, temperature is fixed at the surfaces of the slab
\begin{equation}
\label{app-sol-steady-T-BC}
T(x=0)=T_{\rm L}, \quad T(x=L)=T_{\rm R}.
\end{equation}
Then the solution to Eq.~(\ref{app-convection-steady-thermo-2}) reads
\begin{equation}
\label{app-convection-steady-tT-sol}
\tilde{T} = T_{\rm L}-T_0 + \frac{\delta T}{L}x,
\end{equation}
where $\delta T=T_{\rm R}-T_{\rm L}$. Clearly, it corresponds to a constant thermal gradient along the $x$ direction.

By using solution (\ref{app-convection-steady-tT-sol}), Eqs.~(\ref{app-sol-steady-static-eqs-cont-1}) and (\ref{app-sol-steady-static-eqs-Gauss-1}) can be rewritten as
\begin{equation}
\label{app-convection-steady-mu-eq}
\Delta \tilde{\mu} = q_{\rm TF}^2\tilde{\mu} +4\pi e^2(\partial_{T} n) \tilde{T},
\end{equation}
where $q_{\rm TF} = \sqrt{4\pi e^2(\partial_{\mu} n)}$ is the Thomas-Fermi wave vector and we used
\begin{equation}
\label{app-sol-steady-static-tn-def}
\tilde{n} = (\partial_{\mu} n) \tilde{\mu} +(\partial_{T} n) \tilde{T}.
\end{equation}
The explicit relations for the square of the Thomas-Fermi wave vector are:
\begin{eqnarray}
\label{app-sol-steady-qTF-3D-expl}
\mbox{3D:} \quad q_{\rm TF}^2 &=& \frac{2e^2N_{g}}{v_F^3 \hbar^3} \left(\mu_0^2+\frac{\pi^2T_0^2}{3}\right),\\
\label{app-sol-steady-qTF-2D-expl}
\mbox{2D:} \quad q_{\rm TF}^2 &=& \frac{4e^2N_{g}}{v_F^2 \hbar^2} T_0 \ln{\left(2\cosh{\frac{\mu_0}{2T_0}}\right)}.
\end{eqnarray}
Notice that $q_{\rm TF}$ is calculated for equilibrium values of the chemical potential $\mu_0$ and temperature $T_0$.

The general solution to Eq.~(\ref{app-convection-steady-mu-eq}) is
\begin{equation}
\label{app-sol-steady-static-mu-sol}
\tilde{\mu} = -\frac{(\partial_{T} n)}{(\partial_{\mu} n)}\tilde{T} +C_1e^{q_{\rm TF} x} +C_2e^{-q_{\rm TF} x},
\end{equation}
where $C_1$ and $C_2$ are integration constants.

Having determined the temperature and chemical potential profiles, let us find the electric field $\tilde{\mathbf{E}}=-\bm{\nabla}\tilde{\varphi}$ inside the slab. We assume that the electric potential is fixed at the surfaces of the slab as follows:
\begin{equation}
\label{app-sol-steady-static-phi-BC-1}
\varphi(x=0) = \varphi_{\rm L}, \quad \varphi(x=L ) = \varphi_{\rm R}.
\end{equation}
According to the Gauss law (\ref{app-sol-steady-static-eqs-Gauss-1}), the electric potential $\tilde{\varphi}$ satisfies the following equation:
\begin{equation}
\label{app-sol-steady-static-u=0-nnotn0-5-Gauss}
\Delta \tilde{\varphi} = \frac{1}{e}\left[q_{\rm TF}^2\tilde{\mu} +4\pi e^2(\partial_{T} n) \tilde{T}\right].
\end{equation}
Taking into account Eq.~(\ref{app-sol-steady-static-mu-sol}), one finds
\begin{equation}
\label{app-sol-steady-static-u=0-nnotn0-5-tphi}
\tilde{\varphi} =C_3+ C_4x+ \frac{1}{e}\left(C_1e^{q_{\rm TF} x} +C_2e^{-q_{\rm TF} x}\right),
\end{equation}
where $C_3$ and $C_4$ are constants. From Eqs.~(\ref{app-sol-steady-static-eqs-NS-1}) and (\ref{app-sol-steady-static-nabla-P}), we see that the following equation should be satisfied as well:
\begin{eqnarray}
\label{app-sol-steady-static-eqs-NS-01}
s_0 \bm{\nabla} \tilde{T} +n_0 \bm{\nabla} \tilde{\mu} = en_0 \bm{\nabla}\tilde{\varphi}.
\end{eqnarray}
Equations~(\ref{app-sol-steady-static-phi-BC-1}), (\ref{app-sol-steady-static-u=0-nnotn0-5-tphi}), and (\ref{app-sol-steady-static-eqs-NS-01}) allow us to fix only three out of four integration constants. To determine the fourth one, we employ the condition of the overall charge neutrality of the sample, i.e.,
\begin{equation}
\label{app-sol-steady-static-eqs-neutral}
\int_0^{L} dx\, \tilde{n} = 0.
\end{equation}
Then we obtain
\begin{eqnarray}
\label{app-sol-steady-static-C1}
C_1 &=& -\frac{C_{\delta T}}{1-e^{L q_{\rm TF}}} -\frac{e\delta \varphi}{2(1-e^{L q_{\rm TF}})},\\
\label{app-sol-steady-static-C2}
C_2 &=& -\frac{C_{\delta T}}{1-e^{-L q_{\rm TF}}} -\frac{e\delta \varphi}{2(1-e^{-L q_{\rm TF}})},\\
\label{app-sol-steady-static-C3}
C_3 &=& \frac{C_{\delta T}}{e} +\frac{\varphi_{\rm R}+\varphi_{\rm L}}{2},\\
\label{app-sol-steady-static-C4}
C_4 &=& -\frac{2C_{\delta T}}{eL},
\end{eqnarray}
where
\begin{equation}
\label{app-sol-steady-static-dCT}
C_{\delta T} = \frac{2\pi e^2 \delta T \left[n_0 (\partial_{T} n) -s_0 (\partial_{\mu} n)\right]}{n_0q_{\rm TF}^2}
\end{equation}
and $\delta \varphi=\varphi_{\rm R}-\varphi_{\rm L}$.

The final solutions for the temperature $\tilde{T}$, the chemical potential $\tilde{\mu}$, the electron density $\tilde{n}$, and the electric field $\tilde{\mathbf{E}}=\tilde{E}\hat{\mathbf{x}}$ in the steady state read
\begin{eqnarray}
\label{app-sol-steady-static-tT}
\tilde{T} &=& T_{L}-T_0 +\frac{\delta T}{L} x,\\
\label{app-sol-steady-static-tmu}
\tilde{\mu} &=& -\tilde{T} \frac{(\partial_{T}n)}{(\partial_{\mu}n)} +\frac{\left(2C_{\delta T}+e\delta \varphi\right)}{2} \frac{\sinh{\left[q_{\rm TF}\left(\frac{2x-L}{2}\right)\right]}}{\sinh{\left(\frac{Lq_{\rm TF}}{2}\right)}},\\
\label{app-sol-steady-static-tn}
\tilde{n} &=& \frac{2C_{\delta T}+e\delta\varphi}{2} (\partial_{\mu}n) \frac{\sinh{\left[q_{\rm TF}\left(\frac{2x-L}{2}\right)\right]}}{\sinh{\left(\frac{Lq_{\rm TF}}{2}\right)}},\\
\label{app-sol-steady-static-tEx}
\tilde{E} &=& \frac{2C_{\delta T}}{eL} -\left(2C_{\delta T}+e\delta \varphi\right) \frac{q_{\rm TF}}{2e} \frac{\cosh{\left[q_{\rm TF}\left(\frac{2x-L}{2}\right)\right]}}{\sinh{\left(\frac{Lq_{\rm TF}}{2}\right)}}.
\end{eqnarray}

\begin{figure*}[!ht]
\centering
\subfigure[]{\includegraphics[width=0.3\textwidth]{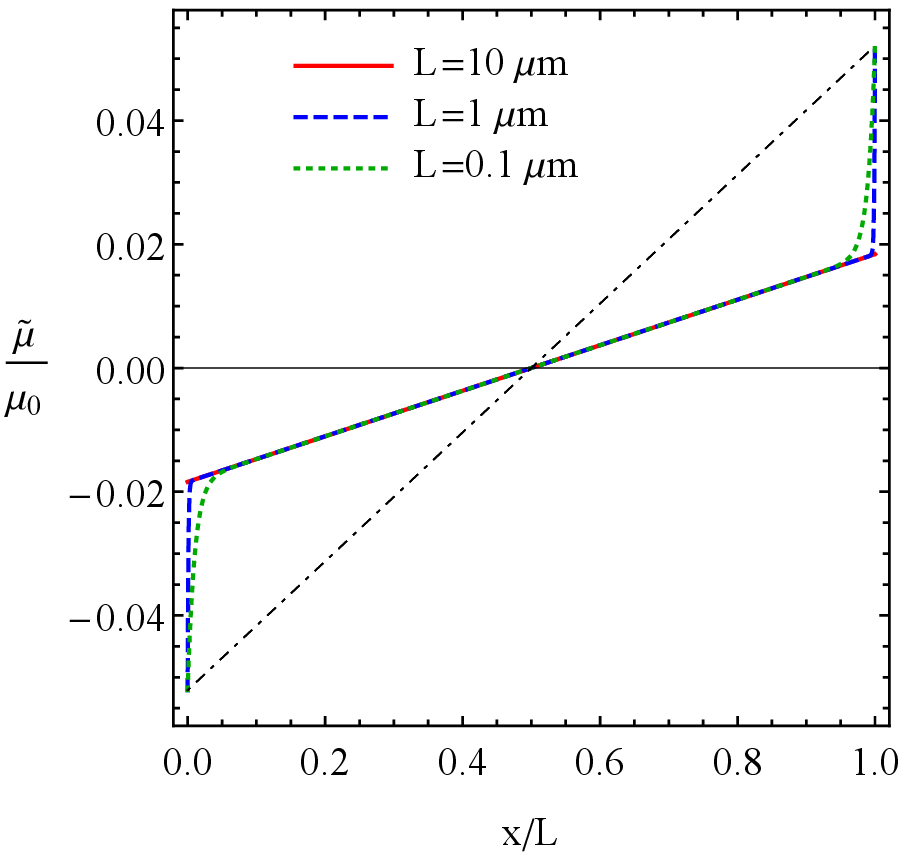}}
\hspace{0.02\textwidth}
\subfigure[]{\includegraphics[width=0.3\textwidth]{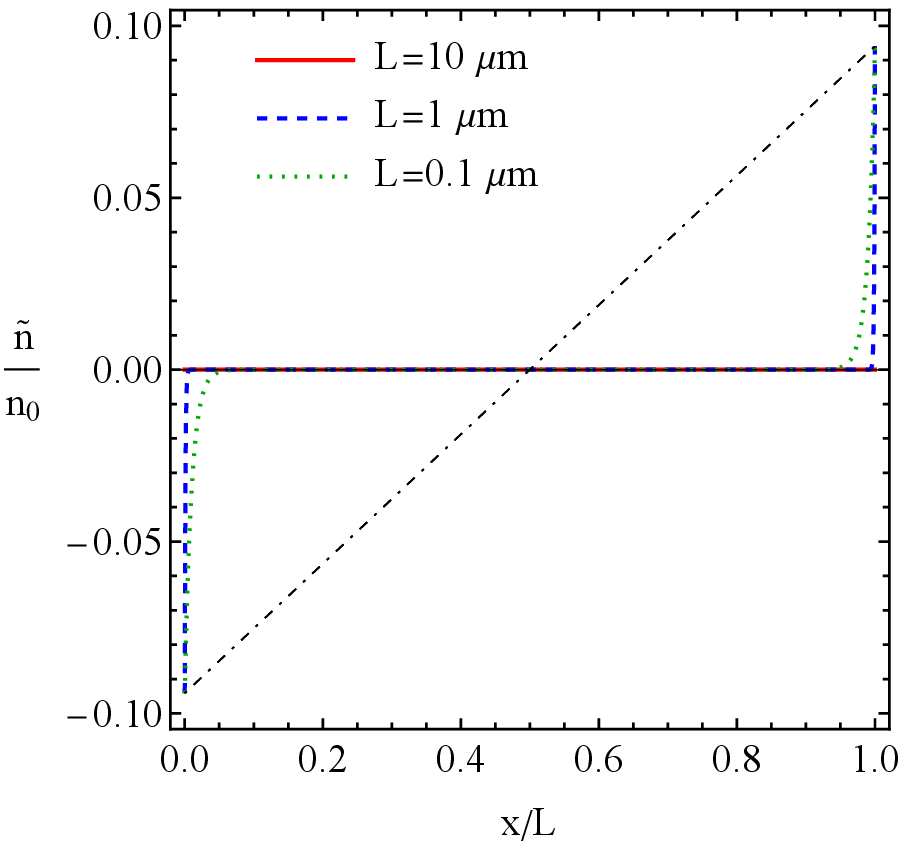}}
\hspace{0.02\textwidth}
\subfigure[]{\includegraphics[width=0.3\textwidth]{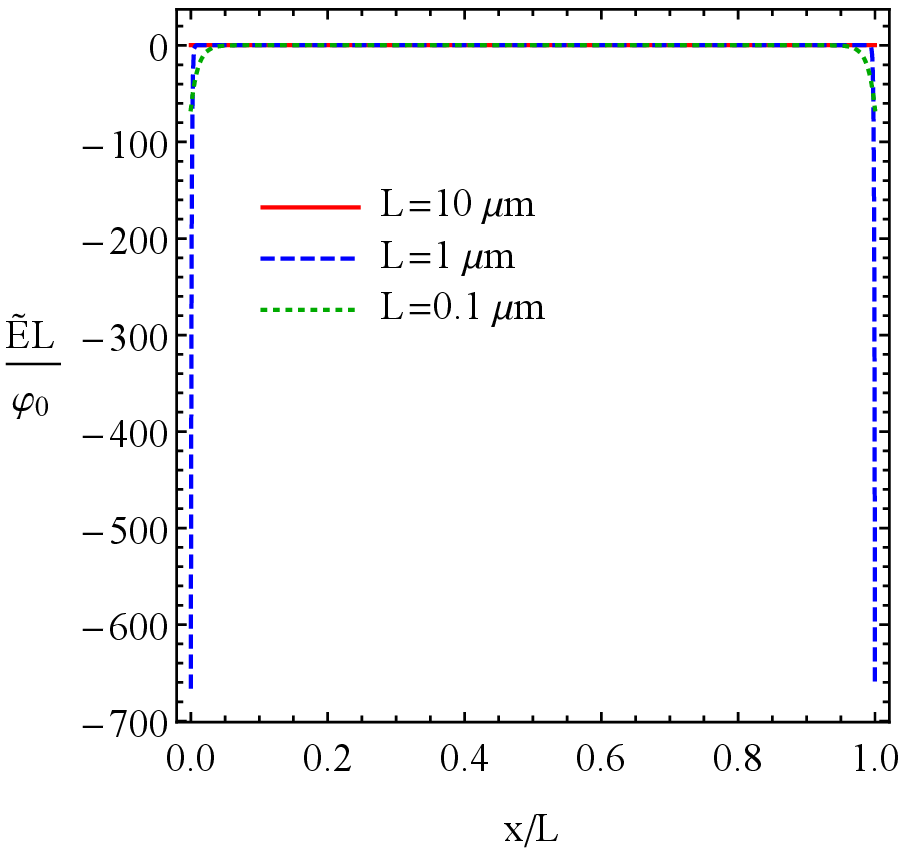}}
\caption{The profiles of the steady-state deviations of the chemical potential $\tilde{\mu}$ (panel (a)), the number density $\tilde{n}$ (panel (b)), and the electric field $\tilde{E}$ (panel (c)). Red solid, blue dashed, and green dotted lines correspond to $L=10~\mu\mbox{m}$, $L=1~\mu\mbox{m}$, and $L=0.1~\mu\mbox{m}$, respectively. The dot-dashed black line corresponds to the model case where the Gauss law is ignored. Numerical parameters are defined in the text.
}
\label{app-fig:sol-steady-static-estimate}
\end{figure*}

It is easy to verify that the chemical potential, the electron density, and the electric field are nonuniform inside the sample. The numerical results for the steady-state solutions (\ref{app-sol-steady-static-tmu}) through (\ref{app-sol-steady-static-tEx}) are presented in Figs.~\ref{app-fig:sol-steady-static-estimate}(a), \ref{app-fig:sol-steady-static-estimate}(b), and \ref{app-fig:sol-steady-static-estimate}(c), respectively. To plot the results, we used the following representative values of parameters: $\mu_0=20~\mbox{meV}$, $T_0=25~\mbox{K}$, and the Fermi velocity for WP$_2$~\cite{Kumar-Felser:2017} $v_F\approx1.4\times10^7~\mbox{cm/s}$. In addition, we set $T_{\rm R}=1.25\times T_0$, $T_{\rm L}=0.75\times T_0$, $\varphi_{\rm L}=0$, and $\varphi_{\rm R}=1~\mbox{mV}$.

\begin{figure*}[!ht]
\centering
\subfigure[]{\includegraphics[height=0.35\textwidth]{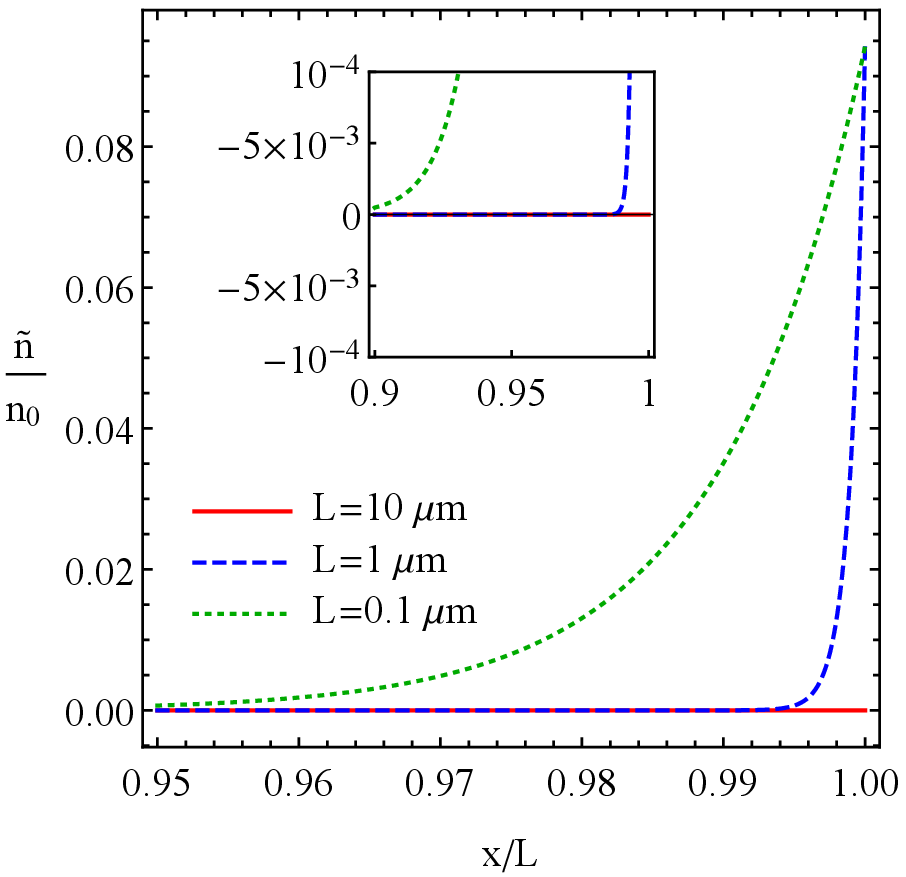}}
\hspace{0.1\textwidth}
\subfigure[]{\includegraphics[height=0.35\textwidth]{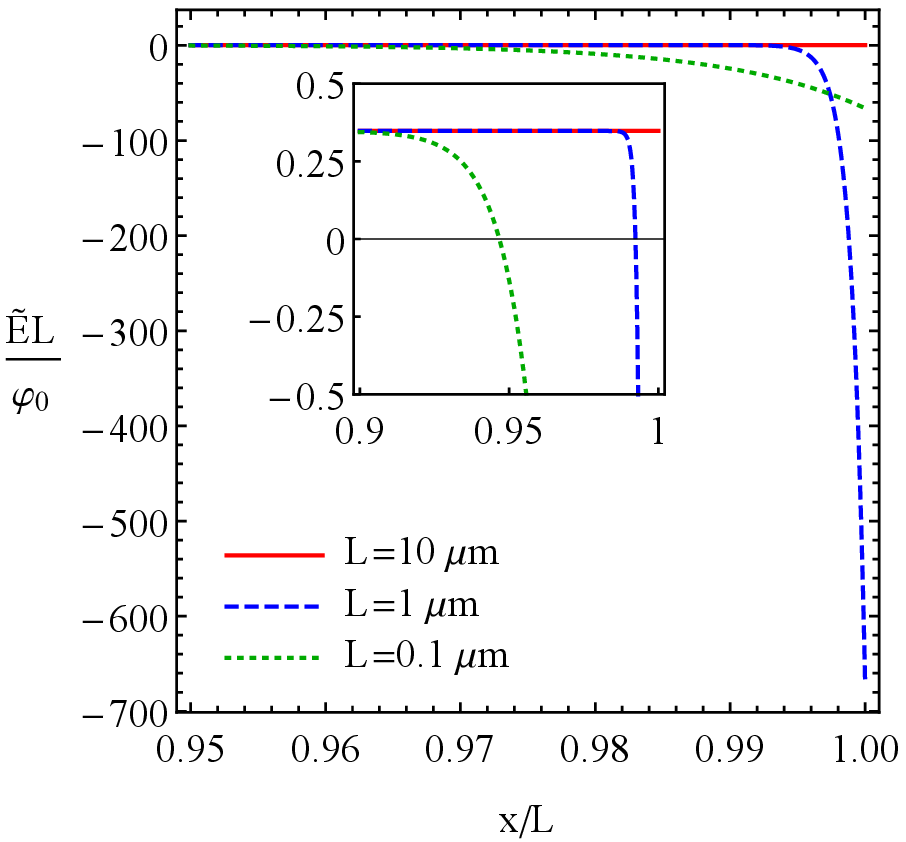}}
\caption{The profiles of the density deviations $\tilde{n}$ (panel (a)) and the electric field $\tilde{E}$ (panel (b)) near one of the surfaces. Insets show zoom-in data. Red solid, blue dashed, and green dotted lines correspond to $L=10~\mu\mbox{m}$, $L=1~\mu\mbox{m}$, and $L=0.1~\mu\mbox{m}$, respectively. Numerical parameters are defined in the text.
}
\label{app-fig:sol-steady-static-estimate-close}
\end{figure*}

As illustrated in Fig.~\ref{app-fig:sol-steady-static-estimate}, the inclusion of the Gauss law strongly affects the solutions for $\tilde{\mu}$ and $\tilde{n}$. First of all, the electron density deviations quickly vanish in the bulk of the slab. This is also confirmed by the dependence of $\tilde{n}$ near one of the surfaces shown in Fig.~\ref{app-fig:sol-steady-static-estimate-close}(a). The electric field $\tilde{\mathbf{E}}$ is screened strongly but not perfectly inside the slab, see Fig.~\ref{app-fig:sol-steady-static-estimate-close}(b). A small nearly uniform value of $\tilde{E}$ survives inside the slab. Its value is determined by the thermoelectric effect, which is quantified by the first term in Eq.~(\ref{app-sol-steady-static-tEx}). For the parameters at hand, we estimate the corresponding field as
\begin{equation}
\label{app-sol-steady-static-tEx-thermo}
\tilde{E} \sim \frac{2C_{\delta T}}{eL} \approx 0.03\, L^{-1}\mbox{[cm]}~\mbox{V/m}.
\end{equation}
Therefore, for the slab width $L=100~\mu\mbox{m}$ to $L=1~\mbox{cm}$, the in-medium electric field ranges from $\tilde{E} \sim 3~\mbox{V/m}$ to $\tilde{E} \sim0.03~\mbox{V/m}$.
These results allow us to approximate the charge density and the electric field by $\tilde{n}\approx0$ and $\tilde{E} \approx const$ in the bulk of the slab. This will be our starting point in determining the convection threshold in the next section.

\section{S III. Convection threshold}
\label{app-sec:Solutions-steady-hydro}

In this section, we investigate the threshold of convective instability of the electron fluid. The steady-state solution given in Eqs.~(\ref{app-sol-steady-static-tT}) through (\ref{app-sol-steady-static-tEx}) will be considered as a background for hydrodynamic flow, i.e.,
\begin{eqnarray}
\label{app-sol-steady-hydro-T-series}
T &=& T_0 + \tilde{T}+T_u,\\
\label{app-sol-steady-hydro-mu-series}
\mu &=& \mu_0 + \tilde{\mu} +\mu_u,\\
\label{app-sol-steady-hydro-E-series}
\mathbf{E} &=& \tilde{\mathbf{E}}+\mathbf{E}_u,
\end{eqnarray}
where the subscript $u$ denotes the deviations related to the fluid flow $\mathbf{u}$.

\subsection{S III.A. 3D Dirac semimetals}
\label{app-sec:Solutions-steady-hydro-3D}

Let us start with the case of 3D Dirac semimetals. As in the case of conventional fluids, we linearize the system of hydrodynamic equations in $\mathbf{u}$, $\mu_u$, and $T_u$. The resulting equations are
\begin{eqnarray}
\label{app-sol-steady-u-Hydro-Weyl-NS-1}
&&\bm{\nabla}P_u -\eta \Delta \mathbf{u} -\left(\zeta+\frac{\eta (d-2)}{d}\right) \bm{\nabla}\left(\bm{\nabla}\cdot\mathbf{u}\right) = -\frac{w_0 \mathbf{u}}{v_F^2\tau} -en_0\mathbf{E}_u -en_u\tilde{\mathbf{E}},\\
\label{app-sol-steady-u-Hydro-Weyl-energy-1}
&&(\mathbf{u}\cdot\bm{\nabla})\tilde{w}+w_0(\bm{\nabla}\cdot\mathbf{u}) =0,\\
\label{app-sol-steady-u-eqs-cont-1}
&&-en_0(\bm{\nabla}\cdot\mathbf{u})+\sigma(\bm{\nabla}\cdot \mathbf{E}_u) +\frac{\sigma}{e}\Delta \left(\mu_u -\frac{\mu_0}{T_0} T_u\right) = 0,\\
\label{app-sol-steady-u-eqs-Gauss-1}
&&\bm{\nabla}\cdot \mathbf{E}_u = -4\pi e n_u.
\end{eqnarray}
In writing the above system, we took into account that the steady-state deviations are much smaller than the equilibrium values. Accordingly, we omitted terms quadratic in the steady-state deviations, e.g., $\tilde{n}\tilde{\mathbf{E}}$. The cross-terms containing both hydrodynamic variables and steady-state deviations are also assumed to be small. However, as we discuss in the main text, there are two notable exceptions from this scheme. Similar to the Oberbeck-Boussinesq approximation for neutral fluids~\cite{Landau:t6-2013,Chandrasekhar:book}, we retain the buoyancy-like term $-en_u\tilde{\mathbf{E}}$ in the Navier-Stokes equation. It plays a crucial role in convection by providing a connection between the fluid velocity and hydrodynamic deviations of temperature and chemical potential. Finally, since convection relies on the compressibility of a fluid, we retained $(\mathbf{u}\cdot\bm{\nabla})\tilde{w}$ in the energy conservation equation. This eventually allows for an advection-like term in the equation for $T_u$.

By using the explicit form of the steady-state solutions from
Sec.~S II, we obtain
\begin{eqnarray}
\label{app-sol-steady-uneq0-2-shorthand-1}
&&(\mathbf{u}\cdot\bm{\nabla})\tilde{w} = u_x (\partial_x \tilde{w}),\\
\label{app-sol-steady-uneq0-2-shorthand-2}
&&\tilde{\mathbf{E}} = \tilde{E} \hat{\mathbf{x}},\\
\label{app-sol-steady-uneq0-2-shorthand-3}
&&(\partial_x \tilde{w}) = (d+1) \left[s_0 -n_0 \frac{(\partial_{T}n)}{(\partial_{\mu}n)}\right] \frac{\delta T}{L}.
\end{eqnarray}
Notice that while we keep the spatial dimension $d$ in the above equations, the system of equations (\ref{app-sol-steady-u-Hydro-Weyl-NS-1}) through (\ref{app-sol-steady-u-eqs-Gauss-1}) will be modified in 2D. We discuss the 2D case in Sec.~S III.B.

By using Eq.~(\ref{app-sol-steady-u-Hydro-Weyl-energy-1}), one can reexpress $\bm{\nabla}\cdot\mathbf{u}$ in terms of $u_x$. Then Eq.~(\ref{app-sol-steady-u-eqs-cont-1}) can be brought in a form resembling the conventional heat transfer equation, i.e.,
\begin{equation}
\label{app-sol-steady-u-eqs-cont-2}
\frac{en_0}{w_0}u_x (\partial_x \tilde{w}) -\frac{\sigma\mu_0}{eT_0} \Delta T_u -4\pi e\sigma n_u +\frac{\sigma}{e}\Delta \mu_u = 0.
\end{equation}
The last two terms on the left-hand side of the above equation clearly distinguish it from the conventional heat transfer equation with an advection term. It is convenient to reexpress $u_x$ in terms of $T_u$ and $\mu_u$, i.e.,
\begin{equation}
\label{app-Solutions-steady-hydro-ux-1}
u_x = -\frac{\sigma w_0}{e^2n_0 (\partial_x \tilde{w})} \left\{ \left(\Delta  -q_{\rm TF}^2\right)\mu_{u}  -\frac{\mu_0}{T_0}\left[\Delta +4\pi e^2 (\partial_{T} n)\frac{T_0}{\mu_0}\right] T_{u}\right\}.
\end{equation}

To determine the relationship between temperature and chemical potential, we calculate the divergence of Eq.~(\ref{app-sol-steady-u-Hydro-Weyl-NS-1}) and obtain
\begin{equation}
\label{app-Solutions-steady-hydro-eps}
\frac{1}{d}\Delta \epsilon_u = -\eta \frac{(\partial_x \tilde{w})}{w_0} \left(\frac{2(d-1)}{d}\Delta -\frac{1}{\lambda_{G}^{2}}\right)u_x +4\pi e^2 n_0 n_u -e\tilde{E} (\partial_x n_u),
\end{equation}
where
\begin{equation}
\label{app-Solutions-steady-hydro-Rtau-def}
\lambda_{G} = \sqrt{\frac{v_F^2\tau \eta}{w_0}} = \sqrt{\tau \eta_{\rm kin}},
\end{equation}
is the Gurzhi length, $\eta$ is the dynamic shear viscosity, and $\eta_{\rm kin} = v_F^2\eta/w_0$ is the kinematic viscosity. The Gurzhi length quantifies the momentum relaxation effects due to the scattering on impurities and phonons. Notice that the first and last terms on the right-hand side in Eq.~(\ref{app-Solutions-steady-hydro-eps}) can be omitted in the perturbation scheme used. Indeed, they contain small gradient terms multiplied by hydrodynamic variables. By neglecting these terms and rewriting $\epsilon_u$ and $n_u$ in terms of the chemical potential $\mu_u$ and temperature $T_u$, we obtain
\begin{equation}
\label{app-Solutions-steady-hydro-eps-1}
n_0\left(\Delta -q_{\rm TF}^2\right)\mu_u + \left[s_0\Delta -4\pi e^2 n_0 (\partial_{T} n)\right]T_u =0.
\end{equation}
Finally, applying curl twice to Eq.~(\ref{app-sol-steady-u-Hydro-Weyl-NS-1}) and taking the $x$-component, one obtains
\begin{equation}
\label{app-Solutions-steady-hydro-rot-rot}
\eta \left(\Delta -\frac{1}{\lambda_{G}^{2}}\right) \left[\Delta +\frac{(\partial_x \tilde{w})}{w_0}\partial_x\right] u_x= e \tilde{E} \left(\Delta-\partial_x^2\right) n_u.
\end{equation}
As before, the term $\propto (\partial_x \tilde{w})$ could be neglected in the above equation leading to
\begin{equation}
\label{app-Solutions-steady-hydro-rot-rot-1}
\left(\Delta -\frac{1}{\lambda_{G}^{2}}\right) \Delta u_x= \frac{e \tilde{E}}{\eta} \left(\Delta-\partial_x^2\right) \left[(\partial_{\mu} n) \mu_u +(\partial_{T} n)T_u\right].
\end{equation}
Equations (\ref{app-Solutions-steady-hydro-ux-1}), (\ref{app-Solutions-steady-hydro-eps-1}), and (\ref{app-Solutions-steady-hydro-rot-rot-1}) form the system of equations for $u_x$, $\mu_u$, and $T_u$.

We search for the solutions of Eqs.~(\ref{app-Solutions-steady-hydro-ux-1}), (\ref{app-Solutions-steady-hydro-eps-1}), and (\ref{app-Solutions-steady-hydro-rot-rot-1}) by employing the plane wave ansatz
\begin{equation}
\label{app-Solutions-steady-hydro-Tu}
T_u = C_T e^{i\mathbf{k}_{\perp}\cdot\mathbf{r}_{\perp}} e^{ik_xx},
\end{equation}
where $\mathbf{k}_{\perp}$ is the wave vector perpendicular to the surface normal. The system has a nontrivial solution if the wave vector satisfies the following characteristic equation:
\begin{equation}
\label{app-Solutions-steady-hydro-char-eq}
\left(k_{\perp}^2+k_x^2\right) \left(k_{\perp}^2+k_x^2+\frac{1}{\lambda_{G}^{2}}\right) \left(k_{\perp}^2+k_x^2 + q_{\rm TF}^2\right) - \frac{k_{\perp}^2}{L^{4}} \mathrm{Ra}=0,
\end{equation}
where
\begin{equation}
\label{app-Solutions-steady-hydro-RQ-def}
\mathrm{Ra} = L^3(d+1)\frac{e^3 n_0 \tilde{E} T_0 \delta T}{\sigma w_0^2\eta (\partial_{\mu}n)}  \left[n_0(\partial_{T}n) -s_0(\partial_{\mu}n)\right]^2
\end{equation}
is the Rayleigh number. Besides the material parameters, it is determined by the electric field and the temperature gradient. The characteristic equation (\ref{app-Solutions-steady-hydro-char-eq}) defines the relation between the wave vector $\mathbf{k}$ and the Rayleigh number. As is clear from the structure of this equation, depending on the sign of $\mathrm{Ra}$, it could have real solutions for wave vector ($\mathrm{Ra}>0$) or only the imaginary ones ($\mathrm{Ra}<0$). The former type of solutions corresponds to periodic convective cells (known as the B\'{e}nard cells in hydrodynamics). Furthermore, a certain threshold value $\mathrm{Ra}_{\rm min}$ should be reached in order to realize convection. To find this threshold, one needs to know the allowed values of $k_x$ which follow from the boundary conditions.

Before defining the boundary conditions, let us present the general expressions of the solutions for the hydrodynamic variables. The solution for $T_u$ reads as
\begin{equation}
\label{app-Solutions-steady-hydro-1}
T_u = \sum_{j=1}^{6}C_T^{(j)} e^{i\mathbf{k}_{\perp}\cdot\mathbf{r}_{\perp}} e^{ik_x^{(j)}x},
\end{equation}
where $k_x^{(j)}$ are six roots of Eq.~(\ref{app-Solutions-steady-hydro-char-eq}) and $C_T^{(j)}$ are constants. The roots are complicated, therefore, we do not present their explicit form here. As to the $x$-component of the fluid velocity, it is given by
\begin{equation}
\label{app-Solutions-steady-hydro-ux-expl}
u_x = -\sum_{j=1}^{6}C_T^{(j)}\frac{\sigma w_0^2}{e^2n_0^2 T_0 (\partial_x \tilde{w})} \left[k^2_{\perp}+\left(k_x^{(j)}\right)^2\right] e^{i\mathbf{k}_{\perp}\cdot\mathbf{r}_{\perp}} e^{ik_x^{(j)}x}
\end{equation}
and does not depend on the sign of $k_x^{(j)}$.

The general solutions for the temperature (\ref{app-Solutions-steady-hydro-1}) and the $x$-component of velocity (\ref{app-Solutions-steady-hydro-ux-expl}) should satisfy the boundary conditions
\begin{eqnarray}
\label{app-Solutions-steady-hydro-BC-1}
T_u(x=0,L)&=&0,\\
\label{app-Solutions-steady-hydro-BC-2}
u_x(x=0,L)&=&0.
\end{eqnarray}
In addition, we employ the free-surface boundary conditions for $\mathbf{u}_{\perp}$, i.e.,
\begin{eqnarray}
\label{app-Solutions-steady-uneq0S-1-BC-freesurf-1}
&&\partial_x u_z(x=0,L)+\partial_z u_x(x=0,L) =0,\\
\label{app-Solutions-steady-uneq0S-1-BC-freesurf-2}
&&\partial_x u_y(x=0,L)+\partial_y u_x(x=0,L) =0.
\end{eqnarray}
By using Eq.~(\ref{app-sol-steady-u-Hydro-Weyl-energy-1}), i.e., $\bm{\nabla}\cdot\mathbf{u}=-(\partial_x\tilde{w})/w_0 u_x$, we rewrite the above relations in a more compact form
\begin{equation}
\label{app-Solutions-steady-uneq0S-1-BC-freesurf-3}
\left(\Delta -2\partial_x^2\right)u_x\approx 0,
\end{equation}
where we used the fact that Eqs.~(\ref{app-Solutions-steady-uneq0S-1-BC-freesurf-1}) and (\ref{app-Solutions-steady-uneq0S-1-BC-freesurf-2}) hold for all $\mathbf{r}_{\perp}$. In addition, we neglected the term $\propto (\partial_x\tilde{w})(\partial_x u_x)$ in Eq.~(\ref{app-Solutions-steady-uneq0S-1-BC-freesurf-3}).
Then one can apply $\partial_z$ and $\partial_y$ to Eqs.~(\ref{app-Solutions-steady-uneq0S-1-BC-freesurf-1}) and (\ref{app-Solutions-steady-uneq0S-1-BC-freesurf-2}), respectively, and solve them in terms of $u_x$.

In the case of the no-slip boundary conditions
\begin{equation}
\label{app-Solutions-steady-uneq0S-1-BC-noslip-1}
u_y(x=0,L)=u_z(x=0,L) =0,
\end{equation}
we obtain
\begin{equation}
\label{app-Solutions-steady-uneq0S-1-BC-noslip-2}
\partial_xu_x(x=0,L)=0.
\end{equation}
While the analysis can be performed for both the free-surface (\ref{app-Solutions-steady-uneq0S-1-BC-freesurf-3}) and no-slip (\ref{app-Solutions-steady-uneq0S-1-BC-noslip-2}) boundary conditions, we consider only the former case. Indeed, the free-surface boundary conditions allow for a much simpler equation for the wave vector $k_x$ whose solutions can be obtained analytically. On the other hand, the case of the no-slip boundary conditions requires numerical calculations (see also Ref.~\cite{Chandrasekhar:book}). While the choice of boundary conditions for the electron velocity results in different values of $k_x$, the final conclusions are qualitatively similar. In addition, by using the arguments in the main text, the use of the precise value of $k_x$ could be avoided altogether.Since the no-slip boundary conditions can be considered as a source of an additional dissipation that only further inhibits convection, we focus only on the free-surface boundary conditions.

By substituting the expressions for $T_u$ and $u_x$ given in Eqs.~(\ref{app-Solutions-steady-hydro-1}) and (\ref{app-Solutions-steady-hydro-ux-expl}) into the boundary conditions (\ref{app-Solutions-steady-hydro-BC-1}), (\ref{app-Solutions-steady-hydro-BC-2}), and (\ref{app-Solutions-steady-uneq0S-1-BC-freesurf-3}), we obtain the following characteristic equation:
\begin{equation}
\label{app-Solutions-steady-uneq0S-char-eq-free-surface}
\sin{\left(L k_x^{(1)}\right)} \sin{\left(L k_x^{(3)}\right)} \sin{\left(L k_x^{(5)}\right)} \left[\left(k_x^{(1)}\right)^2-\left(k_x^{(3)}\right)^2\right] \left[\left(k_x^{(1)}\right)^2-\left(k_x^{(5)}\right)^2\right] \left[\left(k_x^{(3)}\right)^2-\left(k_x^{(3)}\right)^2\right]=0.
\end{equation}
Here, we used the fact that $k_{x}^{(1)}=-k_{x}^{(2)}$, $k_{x}^{(3)}=-k_{x}^{(4)}$, and $k_{x}^{(5)}=-k_{x}^{(6)}$. It is clear that the above equation has a nontrivial solution for $|k_x|=\pi n/L$ with $n=1,2,3,\ldots$.

To make a connection with the literature, let us start with a clean electroneutral system, where $\lambda_{G}\to\infty$ and $q_{\rm TF}=0$. This corresponds to the classical convection of a regular fluid. Then the characteristic equation~(\ref{app-Solutions-steady-hydro-char-eq}) reads
\begin{equation}
\label{Solutions-steady-hydro-RQ-0}
\mathrm{Ra} = L^4\frac{\left(k_{\perp}^2+k_x^2\right)^3}{k_{\perp}^2}.
\end{equation}
The minimum value of the Rayleigh number $\mathrm{Ra}_{\rm min}$ needed to achieve convection is realized for $k_x=\pi/L$ and $k_{\perp,\rm min}=\pi/(\sqrt{2}L)$. It equals
\begin{equation}
\label{Solutions-steady-hydro-RQ-0-min}
\mathrm{Ra}_{\rm min}=\frac{27\pi^4}{4} \approx 657.5.
\end{equation}

Let us now include the effects of the Coulomb forces quantified by $q_{\rm TF}$. For simplicity, we assume that $\lambda_{G}\to\infty$. Then the Rayleigh number $\mathrm{Ra}$ reaches its minimum at
\begin{equation}
\label{app-Solutions-steady-hydro-RQ-kmin}
k_{\perp,\rm min}^2 = \frac{1}{4} \left\{\sqrt{\left[\left(\frac{\pi}{L}\right)^2+q_{\rm TF}^2\right]\left[9\left(\frac{\pi}{L}\right)^2+q_{\rm TF}^2\right]} -q_{\rm TF}^2 -\left(\frac{\pi}{L}\right)^2 \right\}.
\end{equation}
By using this result, the minimal value of the Rayleigh number needed to achieve convection reads
\begin{eqnarray}
\label{app-Solutions-steady-hydro-RQ-qTF-app}
\mathrm{Ra}_{\rm min} &=& \frac{L^4}{8}\left\{ 27 \left(\frac{\pi}{L}\right)^4 -q_{\rm TF}^4 +\left[9\left(\frac{\pi}{L}\right)^2+q_{\rm TF}^2\right] \sqrt{\left[\left(\frac{\pi}{L}\right)^2+q_{\rm TF}^2\right]\left[9\left(\frac{\pi}{L}\right)^2+q_{\rm TF}^2\right]} +18\left(\frac{\pi}{L}\right)^2q_{\rm TF}^2\right\} \nonumber\\
&\stackrel{q_{\rm TF}\to\infty}{\approx}& 4\pi^2L^2 q_{\rm TF}^2.
\end{eqnarray}
As will be evident from the numerical estimates for realistic parameters, $\mathrm{Ra}_{\rm min}$ given in Eq.~(\ref{app-Solutions-steady-hydro-RQ-qTF-app}) is a very large number for semimetals. The case of finite $\lambda_{G}$ and $q_{\rm TF}=0$ can be studied along the same lines. The resulting minimal value of the Rayleigh number is given by Eq.~(\ref{app-Solutions-steady-hydro-RQ-qTF-app}) with $q_{\rm TF}\to \lambda_{G}^{-1}$. In the general case of finite $\lambda_{G}$ and nonvanishing $q_{\rm TF}$, the expression for $\mathrm{Ra}_{\rm min}$ is bulky. Therefore, we do not present it here. Instead, in the main text, we derive the lower bound $\mathrm{Ra}_{0}\leq \mathrm{Ra}_{\rm min}$, which is more than sufficient to conclude that convection is strongly suppressed in Dirac and Weyl semimetals.

As evident from Eq.~(\ref{app-Solutions-steady-hydro-char-eq}), the Coulomb forces and momentum relaxation effects increase the minimal Rayleigh number. To quantify the role of these effects for electron convection, we provide numerical estimates of the minimal Rayleigh number in a few model cases. We notice that, for realistic parameters, $q_{\rm TF}$ is usually much larger than $L^{-1}$. For example, for a 3D relativisticlike energy spectrum, $q_{\rm TF}\approx 9.9\times10^{6}~\mbox{cm}^{-1}$ at $\mu_0=20~\mbox{meV}$, $T_0=25~\mbox{K}$, and the Fermi velocity for WP$_2$~\cite{Kumar-Felser:2017} equals $v_F\approx1.4\times10^7~\mbox{cm/s}$. Therefore, $Lq_{\rm TF}\gg 1$ for typical samples except for very thin films. In addition, although the effect of the Coulomb forces is reduced in a thin film $L\leq q_{\rm TF}^{-1}$, a small thickness of the film is, in fact, unfavorable for convection. Indeed, according to Eq.~(\ref{app-Solutions-steady-hydro-RQ-def}), the actual Rayleigh number $\mathrm{Ra}\propto L^3$. Therefore, one would need much larger fields and/or temperature gradients to reach the convection threshold $\mathrm{Ra}_{\rm min}$. As for the role of the momentum relaxation, we estimate $\lambda_{G}\approx 0.4~\mu\mbox{m}$ at $\tau=0.1~\mbox{ns}$ and $\tau_{ee}\approx0.3~\mbox{ps}$. In this case, $q_{\rm TF}\lambda_{G}\approx382 \gg1$. Therefore, we conclude that the Coulomb forces are a dominant inhibitor of convection in 3D Dirac and Weyl semimetals.

\begin{figure*}[!ht]
\centering
\subfigure[]{\includegraphics[width=0.35\textwidth]{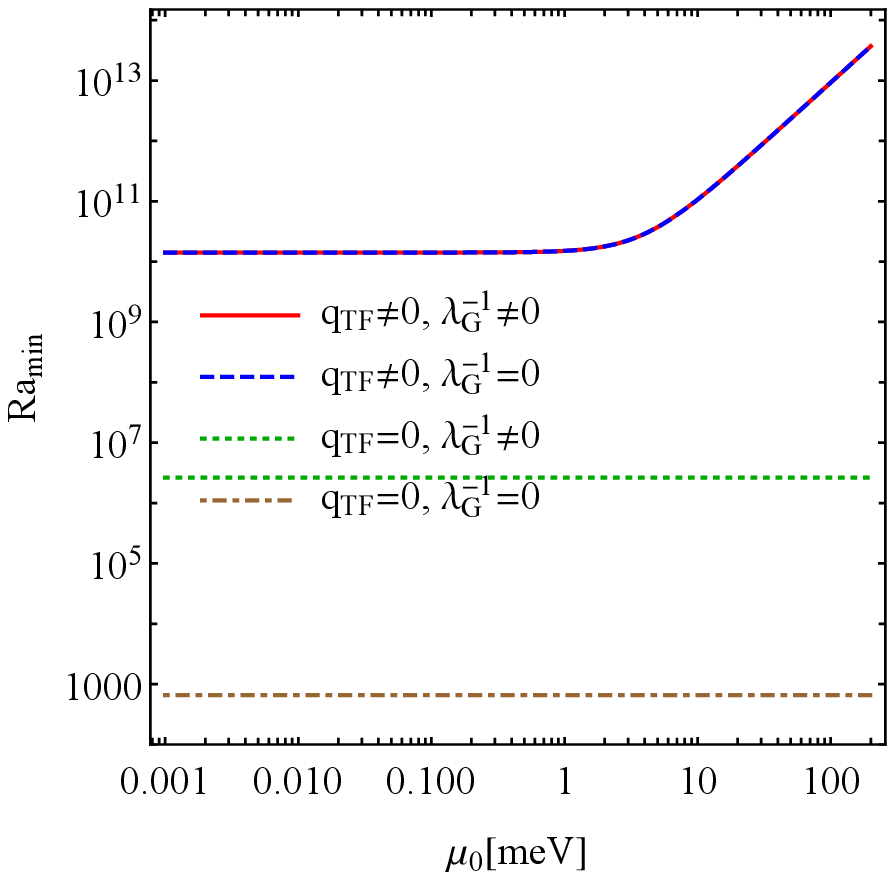}}
\hspace{0.1\textwidth}
\subfigure[]{\includegraphics[width=0.35\textwidth]{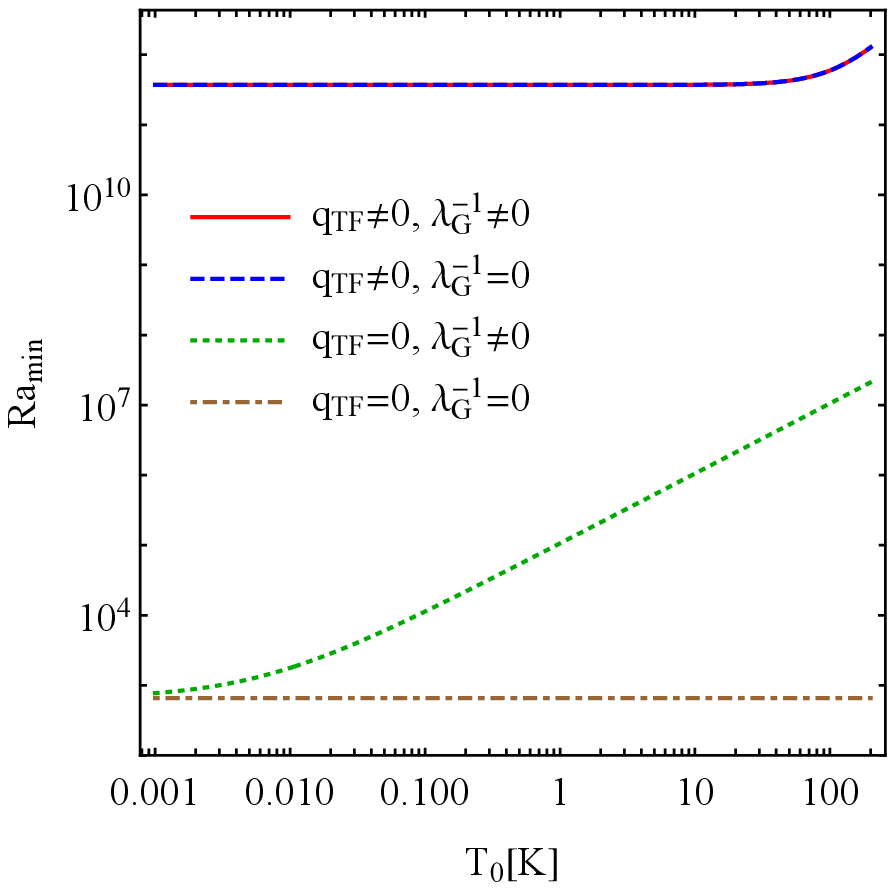}}
\caption{Dependence of the minimal Rayleigh number $\mathrm{Ra}_{\rm min}$ on $\mu_0$ at $T_0=25~\mbox{K}$ (panel (a)) and on $T_0$ at $\mu_0=20~\mbox{meV}$ (panel (b)). Different lines correspond to the following model limits: (i) $q_{\rm TF}\neq0$ and finite $\lambda_{G}$ (red solid lines), (ii) $q_{\rm TF}\neq0$ and $\lambda_{G}\rightarrow\infty$ (blue dashed lines), (iii) $q_{\rm TF}=0$ and finite $\lambda_{G}$ (green dotted lines), (iv) $q_{\rm TF}=0$ and $\lambda_{G}\rightarrow\infty$ (brown dot-dashed lines). In all panels, we set $L=10^{-2}~\mbox{cm}$ and, if appropriate, used values of $q_{\rm TF}$ and $\lambda_{G}$  defined in the main text.
}
\label{fig:app-Ra-min}
\end{figure*}

We provide estimates for the minimal Rayleigh number $\mathrm{Ra}_{\rm min}$ in Fig.~\ref{fig:app-Ra-min} in a few model cases: (i) $q_{\rm TF}\neq0$ and finite $\lambda_{G}$, (ii) $q_{\rm TF}\neq0$ and $\lambda_{G}\rightarrow\infty$, (iii) $q_{\rm TF}=0$ and finite $\lambda_{G}$, (iv) $q_{\rm TF}=0$ and $\lambda_{G}\rightarrow\infty$. As one can see, the minimal Rayleigh number remains much larger than the classical value $27\pi^4/4$ for reasonable values of temperature and chemical potential. This is the case even if the Coulomb forces are omitted. Notice that due to the increase of the electron-electron collision time as $T_0$ decreases (we assumed that $\tau_{ee}\simeq \hbar/T_0$~\cite{Gooth-Felser:2018}), the Gurzhi length rises. This leads to a decrease of $\mathrm{Ra}_{\rm min}$ with temperature at $q_{\rm TF}=0$. We emphasize, however, that the realization of electron hydrodynamics under these conditions is highly unlikely.

\subsection{S III.B. Graphene}
\label{app-sec:graphene}

In this section, we investigate the convection threshold in 2D systems focusing on graphene. We consider a gated sheet of graphene where the gradual channel approximation~\cite{Shur:book,Dyakonov-Shur:1993} (see also Ref.~\cite{Tomadin-Polini:2013} for graphene) can be used instead of the Gauss law (\ref{app-sol-steady-u-eqs-Gauss-1}).
In this case, the hydrodynamically induced electric field $\mathbf{E}_u$ reads as
\begin{equation}
\label{app-Solutions-graphene-capacitance-hydro}
\mathbf{E}_u=  \frac{e}{C} \bm{\nabla} n_u.
\end{equation}
Here, $\delta n$ is the deviation of the charge density from its equilibrium value, $C=\varepsilon/(4\pi L_g)$ is the capacitance per unit area, $\varepsilon$ is the dielectric constant of the substrate, and $L_g$ is the distance to the gate.
The rest of the analysis can be performed along the same line as in
Sec.~S III.A.
In particular, we obtain the following velocity:
\begin{equation}
\label{app-Solutions-graphene-hydro-uz-1}
u_x = -\frac{\sigma w_0}{e^2n_0 (\partial_x \tilde{w})} \Delta \left\{\left[1+\frac{e^2}{C} (\partial_{\mu} n)\right]\mu_u -\frac{\mu_0}{T_0} \left[1-\frac{e^2}{C} (\partial_{T} n) \frac{T_0}{\mu_0}\right]T_u\right\}.
\end{equation}

The analog to Eq.~(\ref{app-Solutions-steady-hydro-eps-1}) reads
\begin{equation}
\label{app-Solutions-graphene-hydro-eps-1}
n_0\left[1 +\frac{e^2}{C} (\partial_{\mu} n)\right]\Delta\mu_u + \left[\frac{(\partial_{T} \epsilon)}{d}\Delta + \frac{e^2}{C} n_0 (\partial_{T} n)\right]\Delta T_u =0.
\end{equation}
Finally, applying curl twice to Eq.~(\ref{app-sol-steady-u-Hydro-Weyl-NS-1}) and taking the $x$-component, one obtains the same relation as in the 3D case, see Eq.~(\ref{app-Solutions-steady-hydro-rot-rot-1}).

By using the plane wave ansatz similar to that in Eq.~(\ref{app-Solutions-steady-hydro-Tu}), where $\mathbf{k}_{\perp}$ and $\mathbf{r}_{\perp}$ are vectors directed along the surface of the graphene ribbon, we obtain the following characteristic equation:
\begin{equation}
\label{app-Solutions-graphene-hydro-char-eq-3}
\left(k_{\perp}^2+k_x^2\right)^2\left(k_{\perp}^2+k_x^2+\frac{1}{\lambda_{G}^{2}}\right)\left(1+Q^2\right) -\frac{k_{\perp}^2}{L^4}\mathrm{Ra} =0,
\end{equation}
where $Q =\sqrt{e^2(\partial_{\mu}n)/C}$. In addition, $\lambda_{G}=v_F\sqrt{\tau \tau_{ee}/4}$ and $\eta =v_F^2\tau_{ee}/4$ in graphene~\cite{Alekseev:2016}. By comparing Eqs.~(\ref{app-Solutions-steady-hydro-char-eq}) and (\ref{app-Solutions-graphene-hydro-char-eq-3}), we notice that the effect of the Coulomb forces quantified by $Q$ is quantitatively different in 2D. However, qualitatively, its role is the same because nonzero $Q$ increases the minimal Rayleigh number needed to achieve convection.

The boundary conditions in 2D have the same form given in Eqs.~(\ref{app-Solutions-steady-hydro-BC-1}), (\ref{app-Solutions-steady-hydro-BC-2}), and (\ref{app-Solutions-steady-uneq0S-1-BC-freesurf-3}) in 3D. They are satisfied at $|k_x|=\pi n/L$. The minimal value of the Rayleigh number is achieved at
\begin{equation}
\label{app-Solutions-graphene-hydro-RQ-kmin}
k_{\perp,\rm min}^2 = \frac{1}{4} \left\{\sqrt{\left[\left(\frac{\pi}{L}\right)^2+\frac{1}{\lambda_{G}^{2}}\right]\left[9\left(\frac{\pi}{L}\right)^2+\frac{1}{\lambda_{G}^{2}}\right]} -\frac{1}{\lambda_{G}^{2}} -\left(\frac{\pi}{L}\right)^2 \right\}
\end{equation}
and reads
\begin{equation}
\label{app-Solutions-graphene-hydro-RQ-qTF-app}
\mathrm{Ra}_{\rm min} = \left(1+Q^2\right)\frac{L^4}{8}\left\{ 27 \left(\frac{\pi}{L}\right)^4 -\frac{1}{\lambda_{G}^{4}} +\left[9\left(\frac{\pi}{L}\right)^2+\frac{1}{\lambda_{G}^{2}}\right] \sqrt{\left[\left(\frac{\pi}{L}\right)^2+\frac{1}{\lambda_{G}^{2}}\right]\left[9\left(\frac{\pi}{L}\right)^2+\frac{1}{\lambda_{G}^{2}}\right]} +18\left(\frac{\pi}{L}\right)^2\frac{1}{\lambda_{G}^{2}}\right\}.
\end{equation}
This expression is similar to that in the 3D case at $q_{\rm TF}=0$ and finite $\lambda_{G}$.

It is clear that Coulomb forces quantified by $Q$ and the momentum relaxation determined by $\lambda_{G}$ increase $\mathrm{Ra}_{\rm min}$. Let us provide numerical estimates for the minimal Rayleigh number and illuminate the role of the Coulomb forces for electron convection in graphene. As we mentioned above, the role of the Coulomb forces is much weaker in 2D than in 3D. Indeed, by using the number density (\ref{app-model-n-2D}), we estimate $Q\approx 6.6$. Here,  we used $v_F=1.1\times10^{8}~\mbox{cm/s}$, $\mu_0=100~\mbox{meV}$, $T_0=100~\mbox{K}$, $L_g=100~\mbox{nm}$, and $\varepsilon=1$. Therefore, the Coulomb forces lead only to an order of magnitude enhancement of $\mathrm{Ra}_{\rm min}$. The momentum relaxation is, however, more important. For example, we estimate $\lambda_{G}\approx 2.6~\mu\mbox{m}$ at $\tau=0.1~\mbox{ns}$ and $\tau_{ee}=1/(\alpha^2) \hbar/T\approx 2.2\times10^{-13}~\mbox{s}$ with $\alpha=e^2/(\hbar v_F)$. Therefore, $L/\lambda_{G}\gtrsim1$ for sufficiently large samples where convection is favorable.

\begin{figure*}[!ht]
\centering
\subfigure[]{\includegraphics[width=0.35\textwidth]{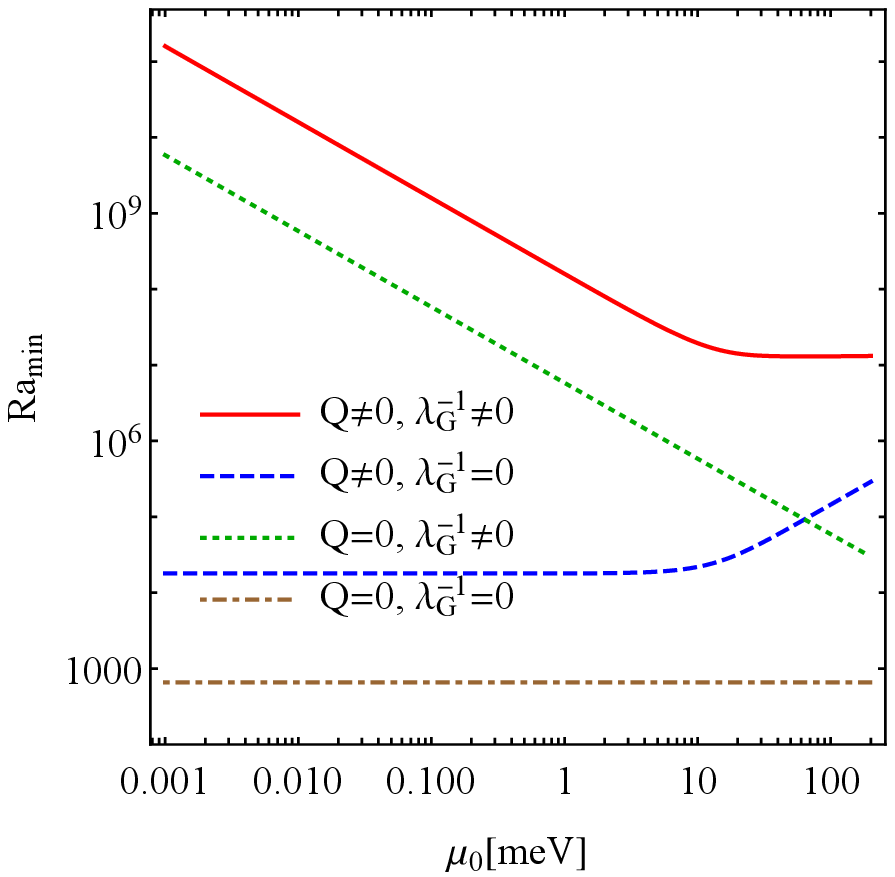}}
\hspace{0.1\textwidth}
\subfigure[]{\includegraphics[width=0.35\textwidth]{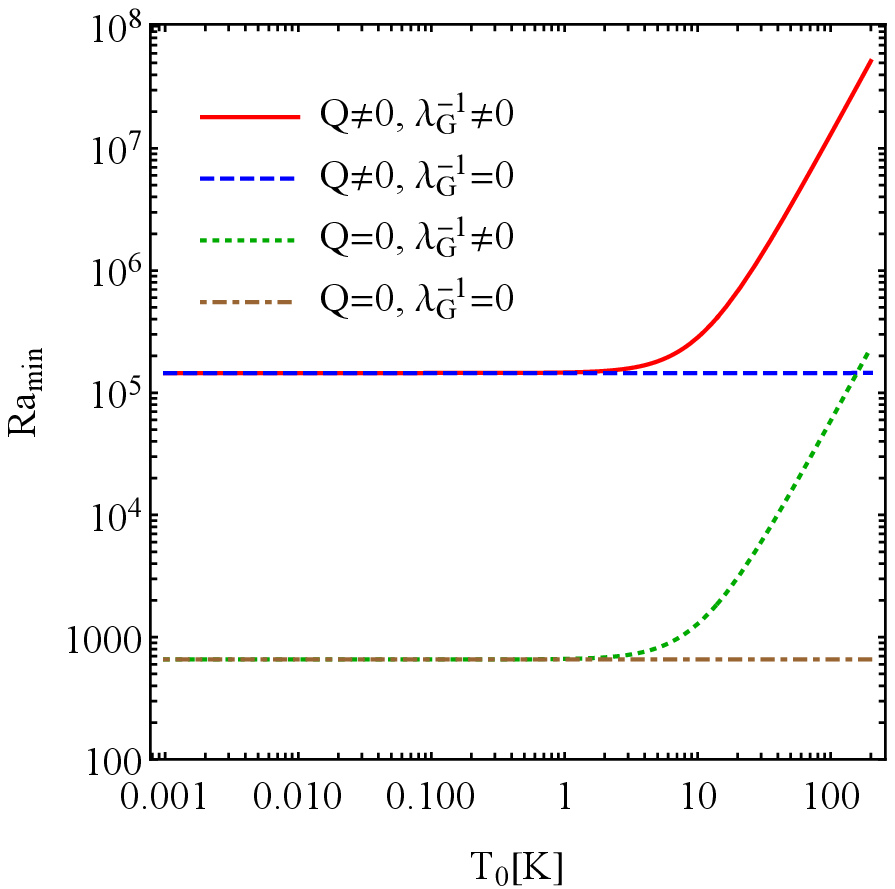}}
\caption{Dependence of the minimal Rayleigh number $\mathrm{Ra}_{\rm min}$ in graphene on $\mu_0$ at $T_0=100~\mbox{K}$ (panel (a)) and on $T_0$ at $\mu_0=100~\mbox{meV}$ (panel (b)). Different lines correspond to the following model limits: (i) $Q\neq0$ and finite $\lambda_{G}$ (red solid lines), (ii) $Q\neq0$ and $\lambda_{G}\rightarrow\infty$ (blue dashed lines), (iii) $Q=0$ and finite $\lambda_{G}$ (green dotted lines), (iv) $Q=0$ and $\lambda_{G}\rightarrow\infty$ (brown dot-dashed lines). In all panels, we set $L=10^{-2}~\mbox{cm}$ and, if appropriate, used values of $Q$ and $\lambda_{G}$ defined in the main text.
}
\label{fig:app-graphene-Ra-min}
\end{figure*}

The minimal Rayleigh number $\mathrm{Ra}_{\rm min}$ is shown in Fig.~\ref{fig:app-graphene-Ra-min} for a few limit cases: (i) $Q\neq0$ and finite $\lambda_{G}$, (ii) $Q\neq0$ and $\lambda_{G}\rightarrow\infty$, (iii) $Q=0$ and finite $\lambda_{G}$, (iv) $Q=0$ and $\lambda_{G}\rightarrow\infty$. As in the 3D case, the minimal Rayleigh number remains much larger than the classical value $27\pi^4/4$ for reasonable values of chemical potential and temperature. However, the effect of the Coulomb forces is no longer dominant. Indeed, the momentum relaxation quantified by $\lambda_{G}$ plays an important role too. Because of the much weaker role of the Coulomb forces, the enhancement of the minimal Rayleigh number is more modest than in 3D. Instead of $8$ to $10$ orders of magnitude difference between $\mathrm{Ra}_{\rm min}$ and the classical value $27\pi^4/4$ in 3D Dirac semimetals, we obtain $2$ to $8$ orders of magnitude difference in graphene.

\bibliography{library-short}